\documentclass[
 reprint,
 superscriptaddress,
 amsmath,amssymb,
 aps
]{revtex4-2}

\usepackage[colorlinks,bookmarks=false,citecolor=blue,linkcolor=magenta,urlcolor=blue]{hyperref}
\usepackage[all]{hypcap}
\usepackage{amsmath,amssymb}
\usepackage{graphicx}
\graphicspath{{figures/}{/}}
\usepackage{dcolumn}

\usepackage{bm}
\usepackage[capitalise,compress]{cleveref}
\crefname{section}{Sec.}{Secs.}
\Crefname{section}{Section}{Sections}

\usepackage{verbatim}
\usepackage{color}
\usepackage{soul}
\usepackage{placeins} 
\usepackage{flafter} 
\usepackage{dsfont}

\usepackage{color}

\newcommand{\tr}{{\text{Tr}}}



\begin{document}

\title{Markovian dissipation can stabilize a (localization) quantum phase transition}

\author{Naushad A. Kamar}
\thanks{\text{These authors contributed equally to this work.}}
\affiliation{Department of Physics and Astronomy, Michigan State University, East Lansing, Michigan 48824, USA}
\affiliation{Institute for Theoretical Physics, University of Amsterdam, Science Park 904, 1098 XH Amsterdam, The Netherlands}
\author{Mostafa Ali}
\thanks{\text{These authors contributed equally to this work.}}
\author{Mohammad Maghrebi}
\affiliation{Department of Physics and Astronomy, Michigan State University, East Lansing, Michigan 48824, USA}

\begin{abstract}
    Quantum phase transitions are a cornerstone of many-body physics at low temperatures but have remained elusive far from equilibrium. Driven open quantum systems---a prominent non-equilibrium platform where coherent dynamics competes with Markovian dissipation from the environment--- often exhibit an effective classical behavior. In this work, we present a nontrivial quantum phase transition that is stabilized, rather than destroyed, by Markovian dissipation. We consider a variant of the paradigmatic spin-boson model where the spin is driven and bosons are subject to Markovian loss proportional to frequency (hence, vanishing at low frequencies). We show that the steady state exhibits a localization phase transition where the spin's dynamics is frozen, to be contrasted with the ground-state transition in the absence of dissipation. Furthermore, this transition occurs when the steady state becomes pure. The latter is not simply a dark state of dissipation but rather emerges from a nontrivial renormalization of the spin dynamics by low-frequency bosonic modes. Our work provides a nontrivial example where \textit{quantumness}, typically reserved for ground states, also emerges in dynamical settings, with potential applications in quantum computation.
\end{abstract}
\maketitle

Driven open quantum systems provide a versatile platform to investigate non-equilibrium many-body physics, and can host new phenomena and exotic states of matter that cannot be realized in their equilibrium counterparts \cite{Diehl_Micheli_Kantian_Kraus_Büchler_Zoller_2008,verstraete_quantum_2009}. Their dynamics describe various quantum simulators such as trapped ions \cite{doi:10.1126/science.aad9958, Schindler_Müller_Nigg_Barreiro_Martinez_Hennrich_Monz_Diehl_Zoller_Blatt_2013}, Rydberg gases \cite{Peyronel_Firstenberg_Liang_Hofferberth_Gorshkov_Pohl_Lukin_Vuletić_2012, Firstenberg_Peyronel_Liang_Gorshkov_Lukin_Vuletić_2013, PhysRevLett.111.113901, PhysRevLett.113.023006}, and superconducting qubits \cite{Houck_Türeci_Koch_2012, PhysRevX.7.011016,PhysRevLett.132.010601}.  Furthermore, they can be efficiently realized in programmable quantum simulators \cite{Ma_Saxberg_Owens_Leung_Lu_Simon_Schuster_2019,doi:10.1126/science.adh9932,PRXQuantum.4.040329, Sierant2022dissipativefloquet,haack2023probing}.

At the same time, dissipation together with the non-equilibrium drive generically leads to a loss of quantum coherence, leading to an effective classical behavior \cite{DallaTorre13,Öztop_2012,PhysRevB.93.014307,PhysRevB.74.245316,PhysRevLett.97.236808, PhysRevX.10.011039,sieberer2023universality}. With a few exceptions \cite{Torre2010,Marino2016,Rota_2019,Znidaric_2010}, quantum phase transitions are generically not expected in driven-dissipative settings. This is because the noise spectrum  typically takes the form $A(\omega)=\gamma|\omega| +\Gamma$ in a representative driven-dissipative setting where both quantum colored noise ($\propto \gamma$) and white noise due to Markovian bath ($\propto\Gamma$)  are present. At low frequencies, Markovian dissipation clearly wins and thus governs the critical behavior. In principle, one may consider non-equilibrium colored noise (vanishing with $\omega$) \cite{Torre2010} or \textit{soft} Markovian dissipation (vanishing with the wavevector $k$) \cite{Marino2016}. However, once the fluctuation-dissipation relation is broken (e.g., when noise has no damping counterpart \cite{Torre2010,DallaTorre_2012}), a constant noise level is generated under renormalization group (RG) leading to white noise, and thereby spoiling quantum fluctuations. This should be contrasted against zero temperature where quantum noise is protected by the fluctuation-dissipation relation. This reasoning applies broadly  to driven-dissipative systems, and has become the basis for a \textit{lore} that these systems do not support robust quantum phase transitions. While seemingly very general, the above arguments are based on perturbative RG analyses where a (semiclassical) field-theory description exists. 

In this work, we consider a driven-dissipative variant of the paradigmatic spin-boson model where a spin is (strongly) coupled to bosons which themselves are subject to loss. In the absence of lossy dynamics, this model exhibits a localization quantum phase transition at zero temperature, a phenomenon that is intimately tied to the low-frequency bosonic modes. Dissipation typically spoils this transition as one generically expects. But, contrary to the lore in the literature, we identify a robust quantum phase transition when low-frequency modes are subject to soft dissipation (that vanishes with their frequency). We present a variational ansatz that predicts a sharp transition, and provide numerical evidence using matrix product states \cite{SCHOLLWOCK201196}. Furthermore, we show that upon approaching the transition, the system becomes increasingly pure, adding further credence to the quantum nature of the transition. While previous works reported an effective classical behavior even in spite of soft dissipation, our model involves strong coupling to a single spin~$\frac{1}{2}$ defying a semiclassical treatment. In contrast, we show that a closely related mean-field version of our model exhibits an effectively classical behavior, with the purity vanishing at the transition.

\textit{Model.---}We consider an open quantum system whose dynamics is governed by the Liouvillian $\cal L$:
\begin{equation}
    \frac{d}{dt} \rho = {\cal L}(\rho)=-i [H,\rho] + \sum_k {\cal D}_{L_k}(\rho)\,,
\end{equation}
where $H$ is the system's Hamiltonian and ${\cal D}$ represents the dissipative dynamics in terms of the Lindblad operators $L_k$ as ${\cal D}_L(\rho)= 2L\rho L^\dagger-\{\rho, L^\dagger L\}$. In an open driven system, the Hamiltonian is given in the rotating frame of the drive within the rotating-wave approximation, hence no explicit time-dependent drive. Nevertheless, detailed balance is explicitly broken and the system approaches a nonequilibrium steady state at long times. 
\begin{figure}
\begin{center}
\includegraphics [scale=1.1]{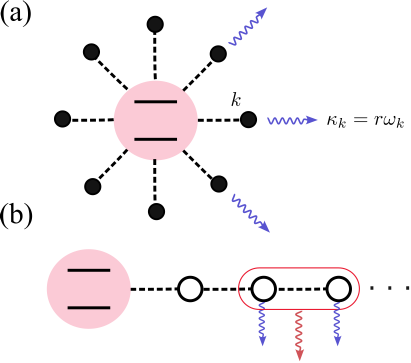}
\end{center} 
\caption{\label{fig:model_dissipation} Schematic picture depicting the spin-boson model with bosons subject to Markovian loss. (a) The top panel shows the spin-boson model where the spin (two-level system) is coupled to bosonic modes (solid circles). Bosons are subject to a mode-dependent loss, $\kappa_k =r\omega_k$, vanishing with the frequency. At large spin-boson coupling, the steady state exhibits a localization phase transition, to be contrasted with the ground-state transition in the absence of dissipation. (b)~The bottom panel depicts the model in a transformed basis where the spin is coupled to a bosonic lattice (open circles). The Markovian loss takes the form of single- and correlated two-site dissipation. 
This model can be realized in superconducting (SC) circuits.
}
\end{figure}
In this work, we consider a single spin coupled to (a continuum of) bosons, a paradigm of quantum impurity models, whose Hamiltonian is given by 
\begin{equation} \label{eq:eq0a}
\begin{split}
H&=-\frac{\Delta}{2} \sigma^x+\sum_k\omega_ka_k^\dagger a_k+\frac{\sigma^z}{2}\sum_k \lambda_k (a_k+a_k^\dagger)\,.
\end{split}
\end{equation}
The spin-bath coupling is characterized by the spectral function
\(J(\omega)=\pi\sum_k \lambda_k^2\delta(\omega-\omega_k)\).
Here, we consider an Ohmic bath with the spectral function 
\begin{equation}\label{eq:spectral1}
\begin{split}
J(\omega)&=2\pi\alpha \omega\Theta(\omega_c-\omega).
\end{split}
\end{equation} 
The parameters $\alpha$ and $\omega_c$ describe the spin-bath coupling and the frequency cutoff of the bath, respectively.
For the  Ohmic bath, this model is extensively studied and is known to exhibit a quantum phase transition to a localized phase where the spin's dynamics is frozen \cite{Leggett_spin_boson_model,weiss2012quantum}. For $\omega_c\gg \Delta$, the phase transition occurs at $\alpha_c= 1+\mathcal{O}(\Delta/\omega_c)$. A qualitative way to understand this transition is to perform a quantum-to-classical mapping where the bosons are integrated out. This leads to a 1D classical Ising model along imaginary time with a long-range coupling of the form $\int d\omega J(\omega) e^{-\omega \tau} \sim \alpha/\tau^2$ where $\tau$ is the distance between classical spins in time. This model is known to undergo a  phase transition due to its long-range coupling \cite{Thouless_1969,Anderson_1971,dyson1971ising,Leggett_spin_boson_model} (otherwise absent in classical 1D systems \cite{peierls1997remarks}). More precisely, the transition is of Beretzinski-Kosterlitz-Thouless (BKT) type with a discontinuous jump of the magnetization \cite{Leggett_spin_boson_model,Anderson_1970,Emery_1974,aizenman1988discontinuity, DeFilippis_2020,Kosterlitz_2016}. Note the importance of the low frequency modes as they are responsible for the long-range coupling in the quantum-to-classical mapping.  Finally, at any finite temperature, the coupling range is limited by the inverse temperature $\beta$, hence no transition. 

In this work, we consider a driven-dissipative variant of the spin-boson model where bosons are not in thermal equilibrium but rather are subject to Markovian dissipation in the form of particle loss
\begin{equation}\label{eq:lindblad}
    L_k = \sqrt{\kappa_k} a_k\,,
\end{equation}
at a rate $\kappa_k$ which we generally allow to be $k$ dependent. 
We thus consider two layers of the bath: the bosons in \cref{eq:eq0a} provide the Ohmic bath while they are also coupled to a Markovian bath.

With Markovian dissipation, the bosonic bath can be still integrated out, but resulting in a different coupling in time. 
Using the Feynman-Vernon formalism \cite{weiss2012quantum}, this coupling is given by (the real and imaginary parts of) the correlator $C(t)\sim\int d\omega J(\omega) e^{-i\omega t - \kappa(\omega)t}$; here, we have defined $\kappa(\omega)=\kappa_k$ such that $\omega_k=\omega$. Let us  first consider a constant dissipation rate, $\kappa_k=\kappa$, independent of $k$. While Markovian loss may appear to project the bosonic bath into its ground state (absent spin coupling), it nevertheless spoils the localization transition. This is because dissipation acts effectively similar to finite temperature as a cutoff on long-range correlations, $C(t)\sim e^{-\kappa t}$. As an alternative, we consider mode-dependent dissipation where low-frequency modes are also slow to decay, hence a soft Markovian dissipation. More precisely, we assume 
\begin{equation}\label{eq:kappa_k}
    \kappa_k = r \omega_k,
\end{equation} 
with $r$ the constant of proportionality; see \cref{fig:model_dissipation}(a) for a schematic figure. In this case, the coupling remains long-ranged, ${C\sim \alpha /[(i+r)t]^2}$, so a phase transition at sufficiently large $\alpha$ is in principle plausible. However, as noted earlier, colored noise at the microscopic level might be spoiled at large scales due to the nontrivial effect of interactions in a non-equilibrium setting. Yet remarkably, we find that a quantum phase transition still emerges, despite strong interactions. 

\textit{Variational ansatz.---}We first present an analytical approach based on a variational ansatz (also referred to as adiabatic renormalization \cite{Leggett_spin_boson_model}) which is perturbative in $\Delta$ but holds for a general value of $\alpha$. We first consider $\Delta=0$ when the Hamiltonian becomes diagonal in the $\sigma^z$ basis, and the steady state takes the form 
\begin{equation}\label{rho_pm}
    \rho_\pm \equiv|\Psi_{\pm}\rangle\langle\Psi_\pm| =|\pm\rangle\langle \pm| \otimes \prod_k|z_{k\pm}\rangle \langle z_{k\pm}|\,.
\end{equation}
Here, $\pm$ denotes the eigenvalues of the $\sigma^z$ operator, and $|z_k\rangle$s denote coherent states for the corresponding mode, $a_k |z_k\rangle= z_k |z_k\rangle$. With the choice of 
\begin{equation}\label{eq:alpha_k}
    z_{k\pm} = \mp \frac{1}{2}\frac{\lambda_k}{\omega_k-i\kappa_k},
\end{equation}
$\rho_\pm$ define steady state solutions of the dynamics. Note that $\rho_\pm$ are pure, hence $|\Psi_\pm\rangle$ are dark states of the Liouvillian. What about coherences, $|\Psi_+\rangle\langle \Psi_-|$? It is easy to see \cite{sm} that the coherence damps out quickly under the dissipative dynamics at the rate $\Gamma_\phi = \sum_k \frac{\lambda_k^2 \kappa_k}{\omega_k^2 + \kappa_k^2} \sim\frac{\alpha r}{1+r^2}\omega_c$, thus an initial coherent superposition quickly becomes a mixture of the states $\rho_\pm$.

Turning on $\Delta$ introduces \textit{quantum fluctuations} which spoil the dark states and generally make the two-level system oscillate between $|\pm\rangle$ states. To treat $\Delta\ne 0$, we use a variational ansatz adapting the Silbey-Harris polaron ansatz \cite{Leggett_spin_boson_model,Silbey_1984,Chin_2011} to steady states of a Liouvillian. To this end, we consider the variational steady state 
\begin{equation}\label{eq:rho_anstaz}
    \rho = \frac{1}{2} \left(|+\rangle \langle+| \otimes |{\cal Z}_+\rangle\langle {\cal Z}_+| +  |-\rangle\langle -| \otimes |{\cal Z}_-\rangle\langle {\cal Z}_-|\right)\,,
\end{equation}
with
\(
    |{{\cal Z}}_\pm\rangle = \prod_k |\zeta_{ k\pm}\rangle 
\)
where $\zeta_{k\pm}\equiv \pm \zeta_k/2$ are variational parameters to be determined. Notice that we have not considered coherences as they die quickly (we always assume $\Delta \ll \omega_c$). To perform the variational calculation, we choose to minimize 
\begin{equation}\label{eq:tr(L^2)}
    \tr (({\cal L}(\rho))^2)\,,
\end{equation}
a non-negative function that only vanishes in the steady state (for a similar choice, see \cite{Cirac_2015}). To compute this quantity, it is convenient to break up the Liouvillian to ${\cal L}={\cal L}_0+{\cal L}_1$ where ${\cal L}_0$ includes the diagonal terms in $\sigma^z$ (i.e., the Liouvillian when $\Delta=0$)
while ${\cal L}_1$ captures the off-diagonal term proportional to $\Delta$. Some algebra then gives
\begin{align}
     \tr (({\cal L}_0(\rho))^2) = \frac{1}{4}|(\omega_k-i\kappa_k)\zeta_{k} +\lambda_k|^2\,,
\end{align}
and 
\begin{align}
    \tr (({\cal L}_1(\rho))^2) 
    = \frac{\Delta^2}{4}(1  -  e^{-\sum_k|\zeta_{k}|^2})\,,
\end{align}
together with $\tr({\cal L}_0(\rho){\cal L}_1(\rho))=0$. Minimizing \cref{eq:tr(L^2)} thus gives the self-consistent equation
\begin{equation}\label{eq:phik_solution}
    \zeta_k =- \frac{\omega_k+i\kappa_k}{\omega_k^2+\kappa_k^2+\Delta_{\rm eff}^2} \lambda_k\,,
\end{equation}
where we have defined
\begin{equation}
    \Delta_{\rm eff}^2 = \Delta^2 \prod_k e^{-|\zeta_k|^2}\,.
\end{equation}
\Cref{eq:phik_solution} reproduces \cref{eq:alpha_k} upon setting $\Delta=0$, and gives a nontrivial result when  $\Delta\ne 0$. The above two equations result in a self-consistent equation for $\Delta_{\rm eff}$ as
\begin{align}\label{eq:tr_L^2_full}
    \Delta_{\rm eff}^2 = \Delta^2 \exp[-\int \frac{d\omega}{\pi } J(\omega)\frac{\omega^2+\kappa(\omega)^2}{(\omega^2+\kappa(\omega)^2+ \Delta_{\rm eff}^2)^2}]\,.
\end{align}
For an Ohmic bath and $\kappa(\omega)=r\omega$, this equation can be solved as
\begin{equation}\label{eq:Delta_eff}
    \Delta_{\rm eff} \propto  \Delta \left(\Delta/\omega_c\right)^{\frac{\tilde\alpha}{1-\tilde\alpha}}\,,\quad \tilde \alpha \equiv \frac{\alpha}{1+r^2} \,.
\end{equation}
It is clear from the above equation that $\Delta_{\rm eff}\to 0$ as $\tilde\alpha\to1$, signaling a localization phase transition where the spin cannot transition between the states $\pm$. This defines a sharp steady-state phase transition at 
\begin{equation}
    \alpha_c = 1+r^2\,,
\end{equation}
in the presence of Markovian dissipation. 
This transition is particularly sensitive to low-frequency bosonic modes: while it occurs for the soft Markovian dissipation where  $\kappa(\omega)\propto \omega$, a constant dissipation rate, for example, merely renormalizes $\Delta$ to a smaller but finite $\Delta_{\rm eff}\ne 0$ and does not lead to any phase transition. 

The above analysis bears close resemblance to the standard spin-boson model (with no Markovian dissipation). 
However, the transitions are fundamentally different: while the latter is a transition in the ground state of the Hamiltonian, the former describes the steady state of a driven-dissipative model. Indeed, the exponent in \cref{eq:Delta_eff}  and the critical value of $\alpha$ are now explicitly dependent on the dissipation (via the parameter $r$). Even in the limit of vanishing dissipation, the steady state is typically far from the ground state \cite{mori_2020}; the fact that $\alpha_c \to 1$ as $r\to0$ coincides with the standard localization transition  is likely an artifact of our ansatz since the coherences are set to zero (not valid in this limit since the damping rate $\Gamma_\phi\to 0$). 

\begin{figure}[t!]
    \centering
    \includegraphics [ scale=0.42]{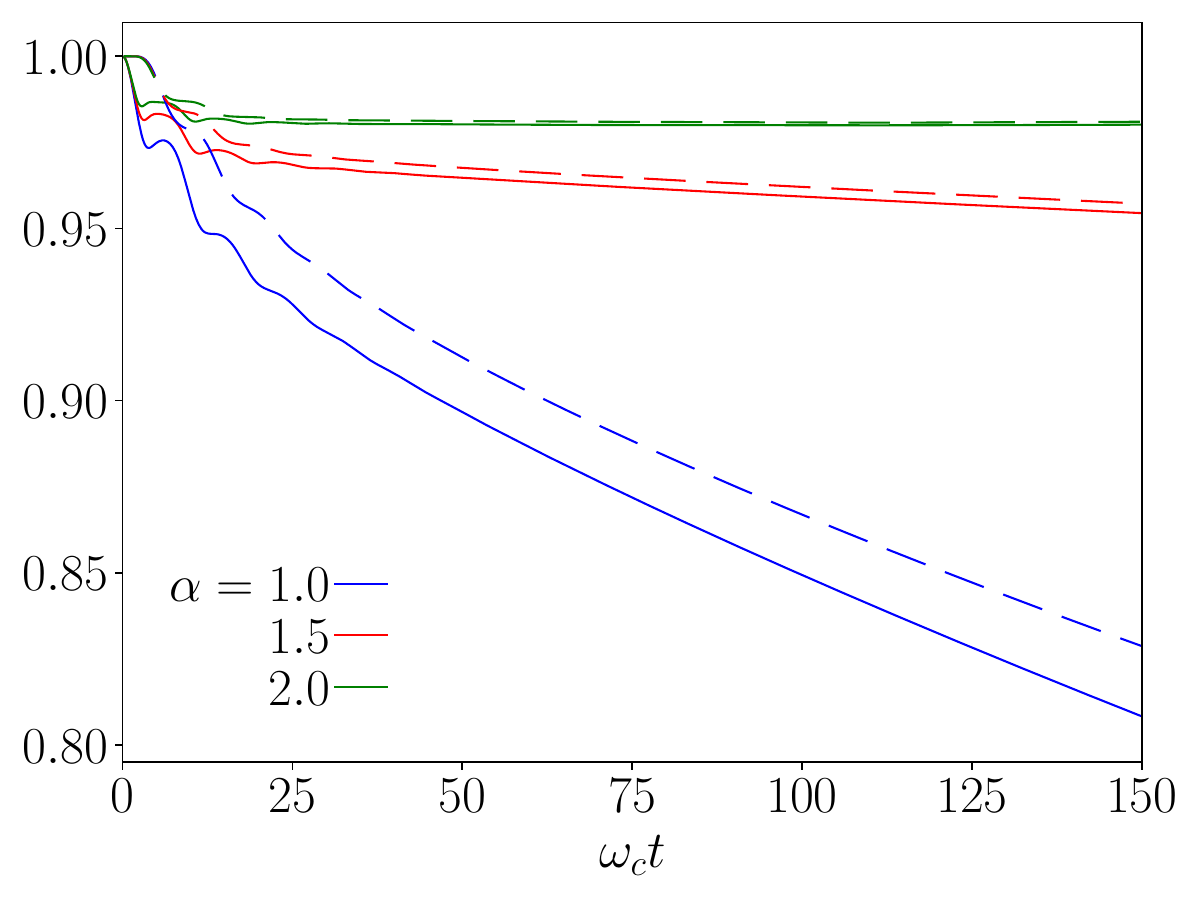}
    \caption{Magnetization $\langle\sigma^z(t)\rangle$ (solid lines) and purity (dashed lines) as a function of time for different values of $\alpha$; we take $\Delta=1,\omega_c=10$, $r=0.5$. 
    For $\alpha= 2$, the magnetization quickly approaches a constant close to unity, signaling a localization phase transition. The purity follows a similar trend as the magnetization, with the spin-boson system remaining almost perfectly pure at the transition. Numerical results are obtained using an optimal bosonic basis with the local bosonic Hilbert space dimension $d_b=12$ and a bosonic chain of size $L=50$.}
    \label{fig:magn+putiry}
\end{figure}

\textit{Numerical simulation.---}The variational ansatz offers a simple, elegant approach to the localization phase transition. Here, we complement this perspective by a numerical analysis using matrix product states (MPS). To this end, it is advantageous to map the spin-boson model to one where the spin is coupled to a semi-infinite bosonic lattice \cite{Plenio_Spin_Boson2010,Chin_2010,Guo_Spin_Boson_Model,Alex_Chin_Spin_Boson2016}; see \cref{fig:model_dissipation}(b). In the absence of dissipation, the Hamiltonian in the new basis is given by 
\begin{equation}\label{eq:chain_mapping}
\begin{split}
H=&-\frac{\Delta}{2} \sigma^x+c_0\sigma^z(b_0+b_0^\dagger)\\
&+\sum_{n=0}^{\infty}\omega_n b_n^\dagger b_n +t_n(b_n^\dagger b_{n+1}+{\rm h.c.}),
\end{split}
\end{equation}
where $c_0=\sqrt{\alpha}\omega_c/2$, describing a \textit{tight-binding} Hamiltonian of the bosonic operators $b_n,b_n^\dagger$ with the on-site energies $\omega_n$ and the hopping amplitudes $t_n$; see \cite{sm}. Similarly, one can transform the dissipation into the new basis; for our choice of dissipation in \cref{eq:lindblad,eq:kappa_k}, we find both single-site and correlated two-site dissipators as 
\begin{equation}\label{eq:L1&2}
    L_{1,n}= \sqrt{\kappa_{1,n}} b_n, \quad L_{2,n}= \sqrt{\kappa_{2,n}} (b_n+b_{n+1}),
\end{equation}
where $\kappa_{1,n}=r (\omega_n - t_n-t_{n-1})$ with $t_{-1} \equiv0$ and $\kappa_{2,n}=r t_{n}$. We note that the single-site dissipation $\kappa_{1,n}$ vanishes as $n\to \infty$ and thus the dissipative dynamics is dominated by the correlated dissipator $L_{2,n}$ at large $n$. In practice, we consider a finite but sufficiently long chain. The resulting lattice model can be realized in superconducting circuits by applying two drive tones on a superconducting qubit while implementing correlated dissipation via coupling to additional lossy qubits \cite{sm}. This type of dissipation has been employed to investigate quantum criticality in different models as well \cite{Marino2016,Begg_2024}.

We are now in a position to numerically simulate the model considered in this work. To this end, we vectorize the density matrix, and utilize the time-evolving block decimation (TEBD) method ~\cite{Vidal_TEBD} combined with the optimal bosonic basis~\cite{Brockt_Optimal_Bosonic_Basis,Stolpp_OBB}. The main challenge in this simulation is however the unbounded local Hilbert space of the bosonic operators. Truncating the dimension of this space to $d_b$ gives a vectorized state of size $d_b^2$, thus limiting the sizes we can simulate. Using an optimal bosonic basis, we can reach only up to $d_b = 12$. This still allows us to observe a phase transition but at relatively large values of $\alpha$, as we report in the SM \cite{sm}; however, this could simply be because the numerics has not fully converged in $d_b$. Here,  we consider a slightly tweaked model where dissipators do not act on site $n=0$; this choice relaxes the required size of $d_b$ (particularly on site $n=0$) to observe the transition. Since the spin couples directly only to $n=0$ [see \cref{eq:chain_mapping}], removing dissipation might seem to drastically alter the model. But recall that the bosonic modes are exactly captured by the correlator $C(t)$. It is straightforward to see that this function would still decay as $1/t^2$ albeit with a slightly different (complex) coefficient \cite{sm}. 
Therefore, we expect that the tweaked model exhibits the same phase transition as the original model, but possibly at a different critical point.

In \cref{fig:magn+putiry}, we present the numerical simulation of the tweaked model starting with the spin in the $|+\rangle$ state and bosons initially in their vacuum state; we take $\Delta=1,\omega_c=10$, $r=0.5$. We see a clear phase transition at $\alpha_c \approx 2$. While our theoretical predication gives $\alpha_c =1+r^2 =1.25$ for the original model with $r=0.5$, the tweaked model displays a transition at a larger value. This could be attributed (beside finite $d_b$) to the prefactor in $|C(t)|\sim 1/t^2$ taking a smaller value upon removing dissipation on site $n=0$ \cite{sm}.

\textit{Nature of phase transition.---}With a localization transition under Markovian dissipation---unlike its zero-temperature analog---it is natural to ask whether the transition is quantum, and if it is a BKT transition. The latter is accompanied with a sudden jump in magnetization at the transition \cite{Leggett_spin_boson_model,Anderson_1970,Emery_1974,aizenman1988discontinuity, DeFilippis_2020,Kosterlitz_2016}. While this appears plausible from our numerics, a detailed numerical analysis is beyond the scope of this work. An analytical treatment via RG \cite{Anderson_1970} is also complicated by open quantum system (Markovian) dynamics. Let us, however, address the quantum nature of the transition. First, we remark that one-dimensional systems [such as \cref{fig:model_dissipation}(b)] generically do not exhibit classical phase transitions \cite{peierls1997remarks}, so the mere existence of a phase transition in our model underscores its quantum nature. To further probe the \textit{quantumness}, we plot the purity of the spin-boson system as a function of time in \cref{fig:magn+putiry}. We observe that, as the transition is approached, the purity follows a trend similar to the magnetization: the system becomes partially mixed below the transition but retains almost perfect purity at the transition. We note that the resulting pure state is not simply a dark state of the Liouvillian, as $\Delta \ne 0$ generally induces nontrivial \textit{tunneling} between the spin states. Yet remarkably, the renormalized tunneling vanishes at the transition, leading to a pure state.

This work provides a nontrivial quantum phase transition stabilized by Markovian dissipation, contrary to the lore that 
driven-dissipative systems become effectively classical in spite of soft dissipation.
It is widely held that driven-dissipative systems become effectively classical, even in the presence of soft dissipation. Our work provides a counterexample. To highlight the contrast, we consider an alternative model where the spin $1/2$ (a hard-core boson) is replaced by a large spin of size $N/2$ (a soft-core boson), resulting in a mean-field Dicke-type model. A superradiant phase transition is thus expected as $N\to\infty$ (regardless of dimensionality). We are rather interested in the \textit{nature} of this phase transition. We find that, even with soft dissipation, fluctuations are effectively classical and diverge upon approaching the phase transition as $\langle S_z^2\rangle \sim 1/\delta$ \cite{sm} in contrast with the quantum scaling, $1/\sqrt{\delta}$, in the ground state (with $S_z$ referring to the total spin, and $\delta$  the distance from the critical point). Furthermore, we show that the steady state's purity quickly vanishes upon approaching the phase transition \cite{sm}. These facts underscore the strongly interacting nature of the spin-boson model, which in turn defies a semiclassical description.

\textit{Conclusions.---}In this work, we have presented a nontrivial example where a strongly interacting, open, driven system undergoes a quantum phase transition. Specifically, we have considered a variant of the spin-boson model with bosons subject to Markovian dissipation. Adapting a variational polaron ansatz to open system dynamics, we have identified a localization phase transition in the steady state---to be contrasted against the ground-state transition in the absence of dissipation. We further support our findings with MPS numerical simulations. Furthermore, we have shown that the spin-boson system becomes pure upon approaching the transition. This is contrasted against a semiclassical variant of the model that becomes effectively classical and increasingly mixed at the transition.  Identifying the precise nature of the phase transition via a systematic RG analysis is a worthwhile avenue for research. An important future direction is to identify what class of driven-dissipative models (e.g., those involving sub-Ohmic baths) exhibit quantum phase transitions. In general, it is highly desired to identify conditions under which quantumness is preserved in dynamical settings far from the quantum ground state, with potential applications to quantum computation.

\textit{Acknowledgments.---}We thank Ron Belyansky, Aash Clerk, Jamir Marino, Sebastian Diehl, and Darrick Chang for useful discussions. This work is supported by the Air Force Office of Scientific Research (AFOSR) under the award number FA9550-20-1-0073, as well as the support from the National Science Foundation under the NSF CAREER Award (DMR-2142866) as well as the NSF grant  PHY-2112893.

\bibliographystyle{apsrev4-2}

\end{document}


\title{Supplemental Material for \\``Markovian dissipation can stabilize a (localization) quantum phase transition''}

\author{Naushad A. Kamar}%
\affiliation{Department of Physics and Astronomy, Michigan State University, East Lansing, Michigan 48824 USA}
\affiliation{Institute for Theoretical Physics, University of Amsterdam, Science Park 904, 1098 XH Amsterdam, The Netherlands}
\author{Mostafa Ali}
\affiliation{Department of Physics and Astronomy, Michigan State University, East Lansing, Michigan 48824 USA}

\author{Mohammad Maghrebi}
\affiliation{Department of Physics and Astronomy, Michigan State University, East Lansing, Michigan 48824 USA}


\maketitle
\setcounter{figure}{0}
\makeatletter
\renewcommand{\thefigure}{S.\@arabic\c@figure}
\setcounter{equation}{0} \makeatletter
\renewcommand{\thesection}{S.\Roman{section}}
\renewcommand \theequation{S.\@arabic\c@equation}
\renewcommand \thetable{S.\@arabic\c@table}

In this Supplemental Material,  we provide additional details on the results stated in the main text. 
In \cref{sec:varitional},  we provide the details on the variational ansatz. In \cref{sec:lattice}, we introduce the mapping of the spin-boson model to a semi-infinite chain, where the spin couples to a one-dimensional bosonic lattice. Specifically, we show that removing dissipation on the first site does not alter the qualitative nature of the bath in \cref{subsec:L_0=0}. In \cref{sec:SC}, we propose an experimental realization of the lattice model using superconducting circuits. In \cref{sec:numerics}, we discuss the details of the MPS numerical methods used in the main text and provide additional numerical results. Finally, in \cref{sec:large_spin}, we show that an effective classical transition emerges in an alternative model where the bosonic bath is coupled to a large spin instead of a single spin $1/2$.

\tableofcontents

\section{Details of Variational Analysis}
\label{sec:varitional}
In this section, we provide the details of the variational analysis, also referred to as adiabatic renormalization \cite{Leggett_spin_boson_model}. It is useful to define the full dynamics in terms of the non-hermitian Hamiltonian 
\begin{equation}
    H_{\rm eff} = -\frac{\Delta}{2} \sigma^x+ \sum_k (\omega_k-i\kappa_k) a^\dagger_k a_k+ \frac{1}{2}\sigma^z\sum_k\lambda_k (a_k+a^\dagger_k)\,.
\end{equation}
The full dynamics is given by
\begin{equation}
    \frac{d}{dt}\rho =-i(H_{\rm eff}\rho -\rho H^\dagger_{\rm eff})+2\sum_k \kappa_k a_k \rho  a^\dagger_k,
\end{equation}
where we denote the last term by the jump term. 

\subsection{Dark state at $\Delta=0$}

We start by considering $\Delta=0$ and the corresponding effective Hamiltonian 
\begin{equation}
    H_{\rm eff, 0} = \sum_k (\omega_k-i\kappa_k) a^\dagger_k a_k+ \frac{1}{2}\sigma^z\sum_k\lambda_k (a_k+a^\dagger_k).
\end{equation}
Given that the dynamics is diagonal in the $\sigma^z$ basis, we find a distinct steady state for each eigenstate ($|\pm\rangle$) of the $\sigma^z$ operator. The corresponding non-Hermitian Hamiltonian in each case is defined as
\begin{equation}
    H_{\rm eff,0 \pm}= \sum_k (\omega_k-i\kappa_k) a^\dagger_k a_k \pm \frac{1}{2}\sum_k\lambda_k (a_k+a^\dagger_k),
\end{equation}
taking the form of a shifted harmonic oscillator for each $k$. Together with the jump terms, it is easy to see that the steady state is pure, and can be written as 
\begin{equation}
    \rho_\pm =|\Psi_{\pm}\rangle\langle\Psi_\pm| =|\pm\rangle\langle \pm| \otimes \prod_k|z_{k\pm}\rangle \langle z_{k\pm}|,
\end{equation}
where $|z_k\rangle$ denotes the coherent state for the corresponding mode: $a_k |z_k\rangle= z_k |z_k\rangle$. To determine the variables $z_{k\pm}$,  the above expression must satisfy the steady state equation:
\begin{align}
    0={\cal L}_0(\rho_\pm)= |\pm\rangle\langle \pm| \otimes \prod_k 
    &\Big (-i \{ (\omega_k-i\kappa_k) a^\dagger_k z_{k\pm}\, {\pm}\, \frac{1}{2}\lambda_k (z_{k\pm}+a^\dagger_k)\} |z_{k\pm}\rangle \langle z_{k\pm}| \nonumber \\
    & +i |z_{k\pm}\rangle \langle z_{k\pm}| \{ (\omega_k+i\kappa_k)  z^*_{k\pm} a_k \,{\pm} \,\frac{1}{2}\lambda_k (a_{k}+z^*_{k\pm})\} \nonumber \\
    & +2\kappa_k |z_{k\pm}|^2 |z_{k\pm}\rangle \langle z_{k\pm}| \Big),
\end{align}
where ${\cal L}_0$ represents the full Liouvillian but with $\Delta=0$. The first two lines on the right hand side of the above equation describe $H_{{\rm eff},0\pm}$ acting from the left and the right, respectively, and the last line is due to the action of the jump term.
One can see that the coefficients of the term $|z_{k\pm}\rangle \langle z_{k\pm}|$ and those with the operators $a_k^\dagger (a_k)$ acting on it from the left (the right) all vanish identically upon choosing
\begin{equation}\label{eq:alpha_k}
    z_{k\pm} = \mp \frac{1}{2}\frac{\lambda_k}{\omega_k-i\kappa_k}.
\end{equation}

Next we consider the coherence:
\begin{equation}
    \rho_{+-} \equiv |\Psi_+\rangle\langle \Psi_-| = |+\rangle\langle -| \otimes \prod_k|z_{k+}\rangle \langle z_{k-}|.
\end{equation}
The Liouvillian acting on the coherence gives 
\begin{align}
    {\cal L}_0(\rho_{+-})= |+\rangle\langle -| \otimes \prod_k 
    &\Big (-i \{\sum_k (\omega_k-i\kappa_k) a^\dagger_k z_{k+}\,{+}\,\frac{1}{2}\lambda_k (z_{k+}+a^\dagger_k)\} |z_{k+}\rangle \langle z_{k-}| \nonumber \\
    & +i |z_{k+}\rangle \langle z_{k-}| \{ (\omega_k+i\kappa_k)  z^*_{k-} a_k \,{-}\, \frac{1}{2}\lambda_k (a_{k}+z^*_{k-})\} \nonumber  \\
    & +2\kappa_k z_{k+}z_{k-}^* |z_{k+}\rangle \langle z_{k-}| \Big).
\end{align}
Substituting $z_{k\pm}$ from \cref{eq:alpha_k}, the coefficients of the term $a^\dagger_k |z_{k+}\rangle \langle z_{k-}|$ and $|z_{k+}\rangle \langle z_{k-}| a_k$ vanish identically, and we find 
\begin{equation}
    {\cal L}_0(\rho_{+-}) = \sum_k(-i \frac{\lambda_k}{2}z_{k+}-i\frac{\lambda_k}{2} z_{k-}^* +2\kappa_k z_{k+}z_{k-}^*) \rho_{+-} = -\sum_k\frac{\lambda_k^2\kappa_k}{ \omega_k^2+\kappa_k^2}\rho_{+-}.
\end{equation}
For a bath with a spectral function $J(\omega)$, the coherence decays at the rate
\begin{equation}\label{eq:G_phi}
    \Gamma_\phi = \int \frac{d\omega}{\pi} J(\omega) \frac{\kappa(\omega)}{ \omega^2+\kappa(\omega)^2}.
\end{equation}
Specifically, for an Ohmic bath and the dissipation rate $\kappa(\omega) =r\omega$, we have
\begin{equation}\label{eq:G_phi_2}
    \Gamma_\phi = \frac{2\alpha r }{1+r^2} \omega_c.
\end{equation}
For a bath with a large cutoff $\omega_c$ and $r$ a constant of order one, the decoherence dies quickly.  

\subsection{Steady state for $\Delta\ne0$}
We now consider $\Delta\ne 0$. We shall adopt a modification of the Silbey-Harris variational approach \cite{Silbey_1984} for the ground state of the spin-boson model (see also Sec. III C of \cite{Leggett_spin_boson_model}) to mixed states of our spin-boson model which is further subject to Markovian dissipation. At a technical level, we adapt the steps in Ref.~\cite{Chin_2011} to our open quantum system. 
To this end, we assume a variational ansatz for the density matrix as
\begin{equation}\label{eq:rho_anstaz}
    \rho = \frac{1}{2}\left( |+\rangle \langle+| \otimes |{\cal Z}_+\rangle\langle {\cal Z}_+| + |-\rangle\langle -| \otimes |{\cal Z}_-\rangle\langle {\cal Z}_-|\right),
\end{equation}
where 
\begin{equation}
    |{\cal Z}_\pm\rangle = \prod_k |\zeta_{\pm k}\rangle ,
\end{equation}
and the variational parameters $\zeta_{\pm k}$ to be determined. The above ansatz describes a mixed state in the presence of Markovian dissipation, similar to the ground-state ansatz $\Psi = \frac{1}{\sqrt2} (|+\rangle \otimes |{\cal Z}_+\rangle + |-\rangle \otimes |{\cal Z}_+\rangle)$ \cite{Leggett_spin_boson_model,Silbey_1984} in the absence of Markovian dissipation. We have not considered a coherence in \cref{eq:rho_anstaz} since any such coherence dies quickly (assuming $\Delta \ll \omega_c$). 

To determine $\zeta_k$, we minimize 
\begin{equation}
    \Tr (({\cal L}(\rho))^2),
\end{equation}
subject to the constraint $\Tr(\rho)=1$. The above function, being a non-negative function, vanishes exactly only for the steady state. In the vectorized notation, this is equivalent to minimizing the Euclidean norm $||\mathbb L |\rho\rrangle||$ for trace-1 density matrices. To this end, we break the full Liouvillian to two parts:
\begin{equation}
    {\cal L}= {\cal L}_0 + {\cal L}_1,
\end{equation}
where ${\cal L}_1$ contains only the term proportional to $\Delta$; all other terms are included in ${\cal L}_0$. We shall see that 
\begin{equation}
    \Tr (({\cal L}(\rho))^2) = \Tr (({\cal L}_0(\rho))^2) + \Tr (({\cal L}_1(\rho))^2).
\end{equation}

Some algebra as before gives
\begin{align}\label{eq:L_line_1}
    {\cal L}_0 (\rho) = 
    \,\,\,\,& \frac{1}{2} |+\rangle \langle+| \otimes \{ \sum_k -i\xi_{k+} a^\dagger_k |{\cal Z}_+\rangle\langle {\cal Z}_+| + i \xi_{k+}^* |{\cal Z}_+\rangle\langle{\cal Z}_+| a_k + \theta_{k+} |{\cal Z}_+\rangle \langle {\cal Z}_+|\}   \nonumber \\
    +&\frac{1}{2} |-\rangle \langle-| \otimes \{ \sum_k -i\xi_{k-} a^\dagger_k |{\cal Z}_-\rangle\langle {\cal Z}_-| + i \xi_{k-}^* |{\cal Z}_-\rangle\langle{\cal Z}_-| a_k + \theta_{k-} |{\cal Z}_-\rangle \langle {\cal Z}_-|\},
\end{align}
where 
\begin{align}\label{eq:xi_k_1}
    \xi_{k\pm}&= (\omega_k-i\kappa_k)\zeta_{k\pm} \,{\pm }\,\frac{\lambda_k}{2}, \\
    \theta_{k\pm} &=     \mp i  \frac{1}{2}\lambda_k \zeta_{k\pm} \,{\pm}\,i\lambda_k \zeta^*_{k\pm}+ 2\kappa_k |\zeta_{k\pm}|^2 .\label{eq:xi_k_2}
\end{align}
Next we compute $\Tr (({\cal L}_0(\rho))^2)$. Let's first consider the contribution due to the first line in \cref{eq:L_line_1}
\begin{align}\label{eq:tr_rho_squared}
    {\rm Tr} \big(&\{ \sum_k -i\xi_{k+} a^\dagger_k |{\cal Z}_+\rangle\langle {\cal Z}_+| + i \xi_{k+}^* |{\cal Z}_+\rangle\langle{\cal Z}_+| a_k + \theta_{k+} |{\cal Z}_+\rangle \langle {\cal Z}_+|\} \nonumber \\
    \times&\{ \sum_{k'} -i\xi_{k'+} a^\dagger_{k'} |{\cal Z}_+\rangle\langle {\cal Z}_+| + i \xi_{k'+}^* |{\cal Z}_+\rangle\langle{\cal Z}_+| a_{k'} + \theta_{k'+} |{\cal Z}_+\rangle \langle {\cal Z}_+|\} 
    \big).
\end{align}
A key simplification occurs upon normal ordering of the bosonic operators, allowing us to substitute 
\begin{equation}\label{eq:subst}
    a_k \to \zeta_{k+}, \qquad a^\dagger_k \to \zeta_{k+}^*,
\end{equation}
but we should also include the contribution due to the nontrivial bosonic commutation relation for $k=k'$. It is easy to see that the contribution from the normal-ordered terms vanishes:
\begin{align}
    -i\xi_{k+} a^\dagger_k |{\cal Z}_+\rangle\langle {\cal Z}_+| + i \xi_{k+}^* |{\cal Z}_+\rangle\langle{\cal Z}_+| a_k + \theta_{k+} |{\cal Z}_+\rangle \langle {\cal Z}_+| \longrightarrow (-i\xi_{k+} \zeta_{k+}^* + i \xi_{k+}^* \zeta_{k+} +\theta_{k+}) |{\cal Z}_+\rangle \langle {\cal Z}_+| =0,
\end{align}
upon the substitution from \cref{eq:xi_k_1,eq:xi_k_2}; this is simply a consequence of $\Tr({\cal L}_0(\rho))=0$. The only nonzero contribution is then due to the nontrivial commutation relations, and is given by
\begin{align}
     \Tr (({\cal L}_0(\rho))^2) = \frac{1}{2} \sum_k|\xi_{k+}|^2 + \frac{1}{2} \sum_k|\xi_{k-}|^2,
\end{align}
where we have used $\langle {\cal Z}_\pm|{\cal Z}_\pm\rangle=1$.

Next we consider the action of ${\cal L}_1$ which contains the Hamiltonian term $ -(\Delta/2)\sigma^x$. The superoperator ${\cal L}_1$ acting on the state $|+\rangle\langle+|$ gives
\begin{equation}
    {\cal L}_1(|+\rangle\langle+| )=i \frac{\Delta}{2} ( \sigma^x |+\rangle\langle+| -  |+\rangle\langle+| \sigma^x) = i \frac{\Delta}{2} (|-\rangle\langle+| -  |+\rangle\langle-|)=- \frac{\Delta}{2}\sigma^y .
\end{equation}
And, similarly, 
\begin{equation}
    {\cal L}_1(|-\rangle\langle-| )= \frac{\Delta}{2}\sigma^y .
\end{equation}
We thus find 
\begin{equation}
    {\cal L}_1(\rho )= -\frac{\Delta}{2}\sigma^y \otimes (\frac{1}{2} |{\cal Z}_+\rangle\langle {\cal Z}_+| - \frac{1}{2}  |{\cal Z}_-\rangle\langle {\cal Z}_-|).
\end{equation}
Also note that 
\(
    \Tr({\cal L}_0(\rho){\cal L}_1(\rho))=0.
\)
We then compute
\begin{align}
    \Tr (({\cal L}_1(\rho))^2) 
    = \frac{\Delta^2}{8}\Tr(I_{\rm spin})  (1 - | \langle {\cal Z}_+|{\cal Z}_-\rangle|^2) 
   = \frac{\Delta^2}{4} \times (1- \prod_k e^{-|\zeta_{k+}-\zeta_{k-}|^2}),
\end{align}
where $I_{\rm spin}$ is the spin's 2$\times$2 identity matrix. We have also used the coherent states' property
\(
    |\langle \beta | \alpha\rangle |^2 = e^{-|\alpha-\beta|^2}
\) in the last equality.
Putting all the terms together, we finally have 
\begin{align}\label{eq:Lrho_squared}
    \Tr (({\cal L}(\rho))^2) = \frac{1}{2} \sum_k|\xi_{k+}|^2 + \frac{1}{2} \sum_k|\xi_{k-}|^2+\frac{\Delta^2}{4} (1- \prod_k e^{-|\zeta_{k+}-\zeta_{k-}|^2}).
\end{align}
We can simplify the above equation by defining  
\begin{equation}
    \zeta_k \equiv \zeta_{k+} - \zeta_{k-}, \qquad \beta_k \equiv\zeta_{k+}+\zeta_{k-}.
\end{equation}
Together with \cref{eq:xi_k_1}, the expression in \cref{eq:Lrho_squared}  takes the form
\begin{align} \label{eq:Lrho_squared_2}
    \Tr (({\cal L}(\rho))^2) = \frac{1}{4} \sum_k \{|(\omega_k -i\kappa_k)\zeta_k +\lambda_k|^2 + |(\omega_k -i\kappa_k)\beta_k|^2 \}+\frac{\Delta^2}{4} (1- \prod_k e^{-|\zeta_{k}|^2}).
\end{align}
The above sum is clearly minimized with $\beta_k=0$;  setting its variation against $\zeta_k$ to zero further yields
\begin{equation}
    (\omega_k+i\kappa_k)\left[(\omega_k -i\kappa_k)\zeta_k +\lambda_k\right] + \Delta^2 \zeta_k \prod_k e^{-|\zeta_k|^2}=0.
\end{equation}
Defining 
\begin{equation}
    \Delta_{\rm eff}^2 = \Delta^2 \prod_k e^{-|\zeta_k|^2},
\end{equation}
we then find
\begin{equation}
    \zeta_k =- \frac{\omega_k+i\kappa_k}{\omega_k^2+\kappa_k^2+\Delta_{\rm eff}^2} \lambda_k.
\end{equation}
The above solution is consistent with \cref{eq:alpha_k} upon setting $\Delta=0$. However, our variational ansatz generalizes our result to a non-trivial $\Delta\ne 0$. The next step is then to self consistently solve for $\Delta_{\rm eff}$ by combining the above two equations:
\begin{align}\label{eq:tr_L^2_full}
    \Delta_{\rm eff}^2 = \Delta^2 \exp\left(-\sum_k \frac{\omega_k^2+\kappa_k^2}{(\omega_k^2+\kappa_k^2+ \Delta_{\rm eff}^2)^2}\lambda_k^2\right) = \Delta^2 \exp\left(-\int \frac{d\omega}{\pi } J(\omega)\frac{\omega^2+\kappa(\omega)^2}{(\omega^2+\kappa(\omega)^2+ \Delta_{\rm eff}^2)^2}\right).
\end{align}
Next considering an Ohmic bath [see \magenta{Eq.~(3)} in the main text] and $\kappa(\omega)=r\omega$, we have 
\begin{align}
    \Delta_{\rm eff}^2 = \Delta^2 \exp\left(-2\alpha (1+r^2) \int_0^{\omega_c} d\omega \frac{\omega^3}{[(1+r^2)\omega^2+ \Delta_{\rm eff}^2]^2}  \right).
\end{align}
Introducing a new notation
\begin{equation}
    \tilde \omega_c \equiv \omega_c^2 (1+r^2), \qquad \tilde\Delta \equiv \Delta^2, \qquad {\tilde\Delta_{\rm eff} \equiv \Delta_{\rm eff}^2}, \qquad \tilde \alpha \equiv \frac{\alpha}{1+r^2},
\end{equation}
the self-consistent equation becomes 
\begin{equation}
    \tilde \Delta_{\rm eff} = \tilde \Delta (1+\tilde\omega_c/\tilde \Delta_{\rm eff})^{-\tilde\alpha} \exp\left[\tilde \alpha(1+ \tilde \Delta_{\rm eff}/\tilde\omega_c)^{-1}\right].
\end{equation}
With the rescaled variables, the above equation becomes identical to that in the standard spin-boson model reported, for example, in Ref.~\cite{Silbey_1984}. Following this reference, we can then identify the solution as 
\begin{equation}
    \tilde \Delta_{\rm eff} = \tilde \Delta \left(\tilde\Delta/ \tilde \omega_c\right)^{\frac{\tilde\alpha}{1-\tilde\alpha}}
\end{equation}
This concludes our derivation of \magenta{Eq.~(15)} in the main text. 

\section{Lattice Representation of the Spin-Boson Model}
\label{sec:lattice}
In this section, we provide a mapping of the spin-boson model to the spin coupled to a bosonic lattice in the presence of dissipation. But we first briefly report the mapping in the absence of dissipation. Our starting point is the Hamiltonian in \magenta{Eq.~(2)} in the main text. Finally, we show that removing dissipation on the first site does not alter the qualitative nature of the bosonic bath in \cref{subsec:L_0=0}.

\subsection{Lattice model without dissipation} 
We first write the Hamiltonian as
\begin{align}
    H= H_{S} + \int_0^{x_{max}}dx g(x) a_x^\dagger a_x +\frac{1}{2}\sigma^z \int_0^{x_{max}} dx h(x) (a_x + a^\dagger_x),
\end{align}
where we have assumed a continuum of bosonic modes which are labeled by $x$; we have also defined $H_S\equiv  -\frac{\Delta}{2}\sigma^x$. The functions $h(x),~g(x)$ define the integral measures, and determine the mode-dependent frequencies and couplings, respectively. The basic idea is to expand the integrals in terms of a set of orthogonal polynomials in a basis designated by the integers $n = 0, 1, \cdots$. We shall not report these polynomials here, and refer the reader to Ref.~\cite{Chin_2010} for details. Using the orthogonality relations, it was shown that the original Hamiltonian maps to a nearest-neighbor interacting model as
\begin{equation}\label{eq:SM4}
\tilde H = -\frac{\Delta}{2} \sigma^x +\frac{\sigma^z}{2} c_0 (b_0 + b_0^\dagger)+\sum_{n=0}^\infty \omega_n b_n^\dagger b_n +\sum_{n=0}^{\infty} t_n( b_n^\dagger b_{n+1} +h.c.),
\end{equation}
where $b(b^\dagger)$ represents the bosonic operator in the new basis. For a bath characterized by the spectral function $J(\omega)=2\pi\alpha\omega^s\omega_c^{1-s}\Theta(\omega-\omega_c)$, the parameters $\omega_n$, $t_n$, and $c_0$ are explicitly given by
\begin{equation}\label{eq:SM5}
\begin{split}
&\omega_n=\frac{\omega_c}{2}\Big(1+\frac{s^2}{(s+2n)(2+s+2n)}\Big),\\&
t_n=\frac{\omega_c(1+n)(1+s+n)}{(s+2+2n)(3+s+3n)}\sqrt{\frac{3+s+2n}{1+s+2n}},\\&
c_0=\sqrt{\frac{\alpha}{2(1+s)}}\omega_c.
\end{split}
\end{equation}
For the Ohmic bath, we set $s=1$.

\subsection{Lattice model with dissipation on all sites}
Next, we consider  bosonic modes  under Markovian dissipation, characterized by the jump operator in \magenta{Eq.~(4)} in the main text. The full Liouvillian dynamics is then given by 
\begin{align}
    \frac{d}{dt} \rho = {\cal L}(\rho) = -i [H, \rho] + \int_0^{x_{max}} dx \kappa(x) (2 a_x \rho a_x^\dagger - a_x^\dagger a_x \rho- \rho a_x^\dagger a_x ).
\end{align}
Notice that the decay rate  $\kappa(x)$ is mode dependent. We consider the dissipation in \magenta{Eq.~(5)} of the main text:
\begin{equation}\label{eq:r}
    \kappa(x) = r g(x),\qquad \mbox{with} \quad r = {\rm const}>0.
\end{equation}
This choice makes the mapping to the spin chain almost identical to Ref.~\cite{Chin_2010}. Next let us consider the superoperator in a vectorized form as\footnote{We choose a basis where $a_x$ is real, i.e., $a_x^* =a_x$; similarly for $\sigma^{x,z}$. Upon vectorization, this allows us to simplify the superoperator $\mathbb L = -i(H\otimes I - I\otimes H^T) + 2L \otimes L^* - L^\dagger L \otimes I - I \otimes L^T L^*$ using $L^*= L, L^T= L^\dagger$, and $H^T= H$.}
\begin{align}
    \mathbb L =& -i (H_{S, u} - H_{S, l}) - i \int_0^{x_{max}}dx g(x) \left[ a_{x,u}^\dagger a_{x,u}- a_{x,l}^\dagger a_{x,l}\right]  \nonumber\\
    &-i \int_0^{x_{max}} dx h(x) \left[\frac{1}{2}\sigma^z_{u} (a_{x,u} + a^\dagger_{x,u})-\frac{1}{2}\sigma^z_{l} (a_{x,l} + a^\dagger_{x,l})\right] \nonumber\\
    &+\int_0^{x_{max}} dx \kappa(x) \left[ 2a_{x,u} a_{x,l}- a^\dagger_{x,u}a_{x,u}-a^\dagger_{x,l}a_{x,l}\right],
\end{align}
where we have used the notation $O_u = O \otimes I$ and $O_l= I\otimes O$. 
With the Markovian dissipation in \cref{eq:r}, we can write the Liouvillian as 
\begin{align}
    \mathbb L =& -i (H_{S, u} - H_{S, l}) - i \int_0^{x_{max}}dx g(x) \left[ (1-ir) a_{x,u}^\dagger a_{x,u}- (1+ir) a_{x,l}^\dagger a_{x,l}\right]  \nonumber \\
    &-i \int_0^{x_{max}} dx h(x) \left[\frac{1}{2}\sigma^z_{u} (a_{x,u} + a^\dagger_{x,u})-\frac{1}{2}\sigma^z_{l} (a_{x,l} + a^\dagger_{x,l})\right] \nonumber \\
    &+2r \int_0^{x_{max}} dx g(x)  a_{x,u} a_{x,l}.
\end{align}
Now the treatment of the first two lines in the above equation is identical to that of \cite{Chin_2010}. The only change is that the \textit{tight-binding} model on the upper/lower leg (denoted by $u,l$) of the ladder should be modified by the corresponding factors of $(1-ir)$ and $(1+ir)$, respectively. The last line of the above equation can be treated similarly, except that it couples the two legs. Writing various terms explicitly, the Liouvillian takes the form
\begin{align}
    \mathbb L = &\quad -i (H_{S, u} - H_{S, l})\nonumber\\& -i\frac{c_0}{2}[\sigma^z_u(b_{0,u}+b_{0,u}^\dagger)-\sigma^z_l(b_{0,l}+b_{0,l}^\dagger)] \nonumber\\
    &-i (1-ir) \sum_{n=0}^\infty \omega_n b^\dagger_{n,u} b_{n,u} + t_n (b^\dagger_{n,u} b_{n+1, u}+ b^\dagger_{n+1,u} b_{n,u}) \nonumber\\
    &+i (1+ir) \sum_{n=0}^\infty \omega_n b^\dagger_{n,l} b_{n,l} + t_n (b^\dagger_{n,l} b_{n+1, l}+ b^\dagger_{n+1,l} b_{n,l}) \nonumber\\
    &+ 2r \sum_{n=0}^\infty \omega_n b_{n,u} b_{n,l} + t_n (b_{n,u} b_{n+1, l}+ b_{n+1,u} b_{n,l}).
    \label{eq:L_first}
\end{align}
The superoperator can be brought into the standard form in terms of Lindblad operators. To see this, we first note that 
\begin{equation}\label{eq:SM10}
b_{n,\alpha} b_{n+1,\beta}+b_{n+1,\alpha} b_{n,\beta}=(b_{n,\alpha}+b_{n+1,\alpha})(b_{n,\beta}+b_{n+1,\beta})-(b_{n,\alpha} b_{n,\beta}+b_{n+1,\alpha} b_{n+1,\beta}),
\end{equation}
where the indices $\alpha,\beta$ are chosen from $\{u,l\}$. One can see that the superoperator can be cast as
\begin{equation}\label{eq:SM11}
\begin{split}
\cal L&=-i (\tilde H_{u} - \tilde H_{l}) +r \sum_{n=0}^\infty \kappa_{1,n} [2b_{n,u} b_{n,l}-b_{n,u}^\dagger b_{n,u}-b_{n,l} ^\dagger b_{n,l}]
\\&+  \sum_{n=0}^{\infty}\kappa_{2,n} [2(b_{n,u}+b_{n+1,u})(b_{n,l}+b_{n+1,l}) -(b_{n,u}^\dagger+b_{n+1,u}^\dagger)(b_{n,u}+b_{n+1,u})-(b_{n,l}^\dagger+b_{n+1,l}^\dagger)(b_{n,l}+b_{n+1,l})],
\end{split}    
\end{equation}
where the Hamiltonian $\tilde H$ is defined in \cref{eq:SM4}, and $\kappa_{1,n}=r (\omega_n-{t}_n-{t}_{n-1})$, $\kappa_{2,n}=t_n$, and $t_{-1}=0$. The above equation depicts the spin-boson model subject to two types of Lindblad operators $L_{1,n}=\sqrt{\kappa_{1,n}}b_n$ and $L_{2,n}=\sqrt{\kappa_{2,n}}(b_n+b_{n+1})$ as defined in \magenta{Eq.~(18)} in the main text.

In practice, we work with a finite bosonic chain of length $L$. In this case, we can truncate the sums in the Liouvillian by setting  $\omega_n,\kappa_{1,n}=0$ for $n>L$ and $t_n,\kappa_{2,n}=0$ for $n\ge L$.

\subsection{Lattice model with dissipation on all sites except $n=0$}
\label{subsec:L_0=0}
Here, we consider a tweaked model where the dissipation is removed on site $n=0$, that is, we set $\kappa_{1,n}=\kappa_{2,n}=0$ for $n=0$. We show that the modified bath has the same qualitative features as the original bath, namely, the bath characteristic function takes the same asymptotic form but with a different coefficient. 
To this end, we calculate the correlator
\begin{equation}
    C(t) = 2 \alpha \omega_c^2\expval{X(t)X(0)},
    \label{eq:two_time_corr_def}
\end{equation}
where the bosonic operator $X(t) = (b_0(t) + b^\dag_0(t))/2$; note that $\alpha$ just appears as an overall prefactor. 
The expectation value is computed in the vacuum state of bosons which is the steady state of the bosonic lattice (absent the coupling to the spin); this is simply the kernel that appears in a Feynman-Vernon influence functional \cite{weiss2012quantum}.
For an Ohmic bath and with Markovian dissipation acting on all sites, this function is explicitly given by (the subscript 0 denoting the model with dissipation on all sites)
\begin{equation}
    \begin{split}
        C_0(t) &= \int_0^{\infty} \frac{d\omega}{2\pi} J(\omega) e^{-i \omega t - \kappa(\omega) t} \\
        &=\frac{i\alpha}{(i+ r)^2} \frac{ 1- e^{-(i+r) \omega_c t }[(i+r) \omega_c t + 1)]}{ t^2}.
    \end{split}
\end{equation}
In the limit $t \gg \omega_c^{-1}$ and for $r>0$, this quantity decays as $1/t^2$.

With dissipation set to zero at site $n=0$, representing the bath in terms of decoupled modes---similar to \magenta{Eq.~(2)} in the main text---is no longer possible. Instead, we employ the \textit{third quantization} techniques in the bosonic operator space \cite{seligman2010}. We consider a bosonic lattice of size $L$, and write the bosonic Hamiltonian  (absent the coupling to the spin) in a compact form as
\begin{equation}
    H_B=\underline{b}^\dag\cdot \mathbf{H}~ \underline{b},
    \label{eq:prosen_Ham}
\end{equation}
where $\underline{b} = (b_0,b_1,...,b_{L-1})^T$ is a vector of bosonic operators. Similarly, the jump operators can be written in a general form as
\begin{subequations}
\label{eq:prosen_jump}
    \begin{align}
        L_{1,n}= \underline{l}_{1,n} \cdot \underline{b},\\
        L_{2,n}= \underline{l}_{2,n} \cdot \underline{b},
    \end{align}
\end{subequations}
with $\underline{l}_{i,n}$ a vector of \textit{c} numbers. 
For our system of interest, the matrix elements of $\mathbf{H}$ are given by
\begin{equation}
    \mathbf{H}_{\alpha,\beta} = \delta_{\alpha,\beta} ~\omega_{\alpha-1} +\delta_{\alpha,\beta-1} ~t_{\alpha-1}+\delta_{\alpha,\beta+1} ~t_{\alpha-2}.
\end{equation}
Similarly, the elements of the vectors $\underline{l}_{1,n}$ and $\underline{l}_{2,n}$ are given by
\begin{subequations}
    \begin{align}
        &(\underline{l}_{1,n})_j = \sqrt{r}\sqrt{ \kappa_{1,n}} \delta_{n,j},\\
        &(\underline{l}_{2,n})_j = \sqrt{r}\sqrt{ \kappa_{2,n}} (\delta_{n,j} + \delta_{n+1,j}).
    \end{align}
\end{subequations}
In the vectorized notation, the bosonic part of the Liouvillian (absent the coupling to the spin) takes the form
\begin{equation}
    {\mathbb L}_B = -i H_B^u + iH_B^l +\sum_{n=0}^{L-1} 2L_{1,n}^u L_{1,n}^{l\dag } - L_{1,n}^{u\dag }L_{1,n}^u - L_{1,n}^l L_{1,n}^{l\dag}+\sum_{n=0}^{L-2} 2L_{2,n}^u L_{2,n}^{ l\dag} - L_{2,n}^{u\dag}L_{2,n}^u - L_{2,n}^l L_{2,n}^{l\dag},
\end{equation}
which can be written in a compact form using \cref{eq:prosen_Ham,eq:prosen_jump} as
\begin{equation}
    \mathbb{L}_B = \underline{b}_0'\cdot (-i\mathbf{H}-\mathbf{M}^*)~ \underline{b}_0 + \underline{b}_1'\cdot (i\mathbf{H}^*- \mathbf{M})~ \underline{b}_1,
\end{equation}
where we have defined the matrix
\begin{equation}
    \mathbf{M} = \sum_{n=0}^{L-1} \underline{l}_{1,n} \otimes \underline{l}_{1,n}^* + \sum_{n=0}^{L-2} \underline{l}_{2,n} \otimes \underline{l}_{2,n}^*,
\end{equation}
and the vectors
\begin{equation}
    \begin{split}
        &\underline{b}_0 =\underline{b}^u,~~~~~~~\underline{b}_0'=\underline{b}^{u*} - \underline{b}^{l*},\\
        &\underline{b}_1 =\underline{b}^{l*},~~~~~\underline{b}_1'=\underline{b}^{ l} - \underline{b}^{u}.
    \end{split}
\end{equation}
We can write the Liouvillian in a more compact form by introducing the $4L \times 4L$ matrix $\mathbf{S}$ as
\begin{equation}
    \mathbb{L}_B = \underline{c} \cdot \mathbf{S}~ \underline{c} - \Tr{\mathbf{M}},
\end{equation}
where $\underline{c} = (\underline{b}, ~ \underline{b}')^T = (\underline{b}_0,~\underline{b}_1,~\underline{b}_0',~\underline{b}_1')$, and $\mathbf{S} = \begin{pmatrix}
    0&-\mathbf{X} \\
    -\mathbf{X}^T &0
\end{pmatrix}$,
with
\begin{equation}
    \mathbf{X} = \frac{1}{2} \begin{pmatrix}
        i \mathbf{H}^* +\mathbf{M} & 0 \\
        0 & - i \mathbf{H} + \mathbf{M}^*
    \end{pmatrix}.
\end{equation}
Assuming the matrix $\mathbf{X}$ is diagonalizable, we can introduce the matrices $\mathbf{P}$ and $\mathbf{\Delta}$ that satisfy
\begin{equation}
    \mathbf{X} \mathbf{P} = \mathbf{P} \mathbf{\Delta},
\end{equation}
where $\mathbf{\Delta} = \text{diag}\{\Delta_1,... , \Delta_{2L}  \}$. The $i$th column of $\mathbf{P}$ represents an eigenvector of $\mathbf{X}$ with the eigenvalue $\Delta_i$. Upon introducing the set of operators $\{\zeta_r\}$ defined by
\begin{subequations}
    \label{eq:3rd_quant_map}
    \begin{align}
        &\underline{b} = (\mathbf{P}^T)^{-1} \underline{\zeta},\\
        &\underline{b}' = \mathbf{P} \underline{\zeta}',
    \end{align}
\end{subequations}
the Liouvillian can be brought to the diagonal form
\begin{equation}
    \mathbb{L}_B = -2 \sum_{r=1}^{2L} \Delta_r \zeta_r' \zeta_r,
\end{equation}
where the operators in the new basis satisfy the condition
\begin{equation}
    \label{eq:vac_ss}
    \zeta_r \ket{\text{NESS}} = \bra{1} \zeta_r' =0,~~~~ \text{for}~r\in \{1,...,2L\}.
\end{equation}

Applying the mapping in \cref{eq:3rd_quant_map} together with \cref{eq:vac_ss}, we can now calculate the correlation function
\begin{equation}
        C(t) = 2\alpha \omega_c^2\expval{X(t)X(0)}
        = \frac{\alpha \omega_c^2}{2} \sum_{r=1}^{2L} \Big[(\mathbf{P}^T)^{-1}_{1,r}+ ( \mathbf{P}^T)^{-1}_{1+L,r}  \Big] \mathbf{P}_{1,r} e^{-2\Delta_r t},
\end{equation}
which is given entirely in terms of the eigenvalues and eigenvectors of the Liouvillian.
Finally, the real and imaginary parts of $C(t)$ characterize the symmetrized correlation function and the response function of the bath; they are respectively given by:
\begin{subequations}
    \begin{align}
        &C^{\text{sym}}(t) =\re C(t)= \alpha \omega_c^2\expval{\acomm{X(t)}{X(0)}},\\
        &\chi(t) = \im C(t) = -i\alpha \omega_c^2 \expval{\comm{X(t)}{X(0)}}.
    \end{align}
\end{subequations}

\begin{figure}[t!]
    \centering
    \includegraphics[width=0.7\linewidth]{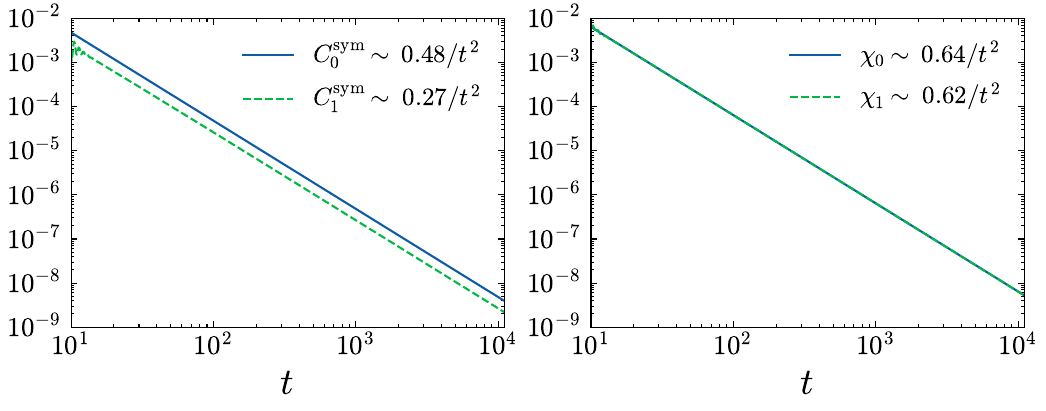}
    \caption{Bath characteristic correlator $C(t)$ as a function of time with the parameters $\alpha = 1,~ r=1/2, ~\omega_c=10$ and the bosonic lattice size of $L=4000$. Left panel: $C^{\text{sym}}_{0/1}(t)= \re C_{0/1}(t)$ represents the symmetrized correlation function. The subscript 0 (1) indicates dissipation on all sites (except site $n=0$). Right panel: $\chi_{0/1}(t)= \im C_{0/1}(t)$ denotes the response function. All the correlators fall off as $a/t^2$ at long times albeit with slightly different coefficients in the two cases.}
    \label{fig:two_time_corr}
\end{figure}

Turning off dissipation on the first site does not alter the asymptotic form of the correlator $C(t)\propto \alpha/t^2$ as $t \to \infty$, except for an overall prefactor. The real and imaginary parts of the correlator $C(t)$ and the asymptotic fits at long times are presented in \cref{fig:two_time_corr}.

\section{Quantum Simulation with Superconducting Circuits}
\label{sec:SC}

\begin{figure}[b!]
    \centering
    \includegraphics[width=0.35\linewidth]{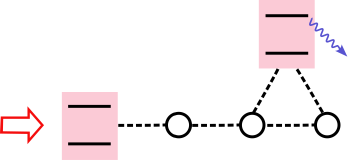}
    \caption{Schematic setup realizing the spin-boson model subject to correlated dissipation. The qubit on the left represents the spin, and the open circles the bosonic sites ($n=0,1,2$ from left to right). The red arrow represents two drive tones which lead to a Rabi-type coupling between the spin and the $n=0$ boson. The right qubit is subject to dissipation, and, upon adiabatic elimination, gives rise to correlated dissipation on $n=1,2$.}
    \label{fig:q_sim}
\end{figure}

Here, we provide a practical scheme to realize the driven-dissipative model considered in the main text. More precisely, we consider  the lattice model while turning off the dissipation on site $n=0$; see \cref{subsec:L_0=0}. For simplicity, we take a lattice of three bosonic sites ($n=0,1,2$); the generalization to many lattice sites is straightforward. The spin (or a qubit, denoted by q1) is coupled to the bosonic site at $n=0$, while a second qubit (denoted by q2) is coupled to sites $n=1,2$ to realize the correlated dissipation; see \cref{fig:q_sim}. 
The basic idea is to combine the quantum simulation of the Rabi model in Refs.~\cite{Ballester_2012,braumuller2017analog} together with the implementation of correlated dissipation in \cite{Marino_2016_PRB,Marcos_2012}. In this scheme, the qubits can, for example, be superconducting qubits coupled to microwave resonator modes.

The Hamiltonian coupling the qubits to bosonic modes is given by 
\begin{align}
    H(t) = & \frac{\omega_{q1}}{2} \sigma^z + \omega_0 b_0^\dagger b_0 -g (\sigma^+ b_0+ \sigma^- b_0^\dagger) -\Omega_1 (e^{i\omega_{p1} t}\sigma^- + {\rm H.c.}) -\Omega_2 (e^{i\omega_{p2} t}\sigma^- + {\rm H.c.}) + \nonumber \\
    &  J_{01} (b_0 b_1^\dagger +{\rm H.c.}) + J_{12} (b_1 b_2^\dagger +{\rm H.c.}) + \omega_1 b_1^\dagger b_1+ \omega_2 b_2^\dagger b_2 + \frac{{\omega_{q2}}}{2}\tau^z + \Omega [(b_1 + b_2) \tau^+ + {\rm H.c.}],
\end{align}
where $\sigma^{\pm},\tau^{\pm}$ represent the raising/lowering spin operators corresponding to the first  and second qubits, respectively. A rotating wave approximation (RWA) is used to cast the coupling between the first qubit and the bosonic mode $b_0$ in the form of the Jaynes-Cummings model.
Following \cite{Ballester_2012}, we also consider two transversal microwave Rabi drive tones $\omega_{p1,p2}$ coupling to the first qubit with amplitudes $\Omega_{1,2}$, respectively, assuming that the RWA applies to the driving term. We also assume that $\omega_{q1,q2, 0,1,2,p1,p2}$ are large compared to all the other frequencies. Finally, the second qubit is coupled to the bosonic modes $b_{1,2}$ with an amplitude $\Omega$ (again assuming RWA) and is also subject to decay at the rate $\Gamma_{q2}$ corresponding to the Lindblad operator $L= \sqrt{\Gamma_{q2}}\tau^-$. Assuming $\Omega_1 \gg \Omega_2$, we can cast the Hamiltonian in the reference frame rotating with the frequency of the driving $\omega_1$. Furthermore, assuming the condition $\omega_{p1} - \omega_{p2}=2\Omega_1$, we can use the RWA to ignore fast oscillations of the order $\Omega_1$ to arrive at the Hamiltonian
\begin{align}
    H_{1} = \,\,\,\,&\frac{\Omega_2}{2}\sigma^z + \delta\omega_{0} b_0^\dagger b_0 - \frac{g}{2} \sigma^x (b_0 + b_0^\dagger)  \nonumber \\
    +& J_{01} (b_0 b_1^\dagger +{\rm H.c.}) + J_{12} (b_1 b_2^\dagger +{\rm H.c.}) + \delta\omega_1 b_1^\dagger b_1+ \delta\omega_2 b_2^\dagger b_2 + \frac{\delta\omega_{q2}}{2}\tau^z + \Omega [(b_1 + b_2) \tau^+ + {\rm H.c.}],
\end{align}
where the frequencies $\delta \omega_i= \omega_i - \omega_{1p}$. Here, we assume $ \Omega_1 $ is large not only compared to $ \Omega_2 $ but also to the couplings $ J_{01}, J_{12}, \Omega $, allowing us to apply the arguments of Ref.~\cite{Ballester_2012} in the presence of additional bosonic modes and the second qubit. Next we assume that the second qubit dynamics is fast, that is, one or both $\delta \omega_{q2}, \Gamma_{q2}$ are large compared to all the remaining frequencies. This allows us to eliminate the second qubit using, for example, the formalism in \cite{Reiter_2012} 
resulting in the effective Hamiltonian 
\begin{align}
 H_{\rm eff} = \,\,\,\,&\frac{\Omega_2}{2}\sigma^z + \delta\omega_{0} b_0^\dagger b_0 - \frac{g}{2} \sigma^x (b_0 + b_0^\dagger) +  \nonumber\\
   + & J_{01} (b_0 b_1^\dagger +{\rm H.c.}) + J_{12} (b_1 b_2^\dagger +{\rm H.c.}) + \delta\omega_1 b_1^\dagger b_1+ \delta\omega_2 b_2^\dagger b_2  + \Delta (b_1^\dagger + b_2^\dagger )(b_1 + b_2 ) ,
\end{align}
where the last term arises from the Lamb shift. Additionally, the system  will be subject to dissipation via the Lindblad operator
\begin{equation}
    L_{\rm eff} = \sqrt{\Gamma} (b_1+b_2).
\end{equation}
In the above equations, we have defined
\begin{equation}
    \Delta = \frac{\Omega^2 \delta \omega_{q2}}{\Gamma_{q2}^2+ \delta\omega_{q2}^2}, \qquad \Gamma = \frac{\Omega^2 \Gamma_{q2}}{\Gamma_{q2}^2+ \delta\omega_{q2}^2}.
\end{equation}
By appropriately tuning the parameters in the above equation, we recover the Hamiltonian in \magenta{Eq.~(17)} in the main text with correlated dissipation acting on sites $n=1,2$. Including single-site dissipation  on the sites $n=1,2$ is trivial but also not essential for realizing the phase transition. 

\section{Numerical Methods and Results}
\label{sec:numerics}

In this section, we briefly present the numerical methods employed in this work. Specifically, we vectorize the density matrix and represent it as a matrix product state (MPS) \cite{SCHOLLWOCK201196}. The time evolution is then carried out using the time-evolving block decimation (TEBD)~\cite{Vidal_TEBD} algorithm, combined with an optimal bosonic basis~\cite{Brockt_Optimal_Bosonic_Basis,Stolpp_OBB} to efficiently capture the  dynamics.

\subsection{Vectorization of density matrix}

Vectorization of the density matrix is achieved by \emph{gluing} the bra and ket indices in a single vector space. More precisely, for a density matrix \(\rho = \sum_{i,j} \rho_{ij} |i\rangle\langle j|\), the vectorized form is defined as
\begin{equation}
|\rho\rangle\rangle = \sum_{i,j} \rho_{ij} |i\rangle \otimes |j\rangle.
\end{equation}
Using this representation, the density matrix can be straightforwardly expressed as a matrix product state:
\begin{equation}
|\rho\rangle\rangle = \sum_{\{s'\}, \{s''\}} W[0]^{s_0's_0''}W[1]^{s_1's_1''} W[2]^{s_2's_2''} \cdots W[L]^{s_L's_L''} 
|s_0's_0'',s_1's_1'', s_2's_2'', \ldots, s_L's_L''\rangle,
\end{equation}
where \( W[n]^{s_n's_n''} \) are the local tensors of the MPS, and \( s_n' \) and \( s_n'' \) denote the local Hilbert space indices corresponding to the bra and ket components at site \( n \). We define the combined index \( s_n \equiv (s_n', s_n'') \), which spans a local Hilbert space of dimension \( d_b^2 \) for \( n > 0 \) (bosonic sites), and \( d_s^2 \) for \( n = 0 \) (spin), where \( d_b \) and \( d_s \) represent the local Hilbert space dimensions of  bosons and the spin, respectively.\footnote{In this section, boson site indices run from $n=1$ to $L$.} Finally, we can express the density matrix in the MPS form as

\begin{equation}
|\rho\rangle\rangle = \sum_{s} A[0]^{s_0}A[1]^{s_1} A[2]^{s_2} \cdots A[L]^{s_L} 
|s_0,s_1, s_2, \ldots, s_L\rangle,
\end{equation}
where $A[n]^{s_n}=W[n]^{s_n's_n''}$ represents the three-dimensional tensor of size $\chi\times\chi\times d_b^2$ for $n>0$ with $\chi$ the MPS bond dimension.
\begin{figure}[h!]
\begin{center}
\includegraphics [ scale=0.4]
{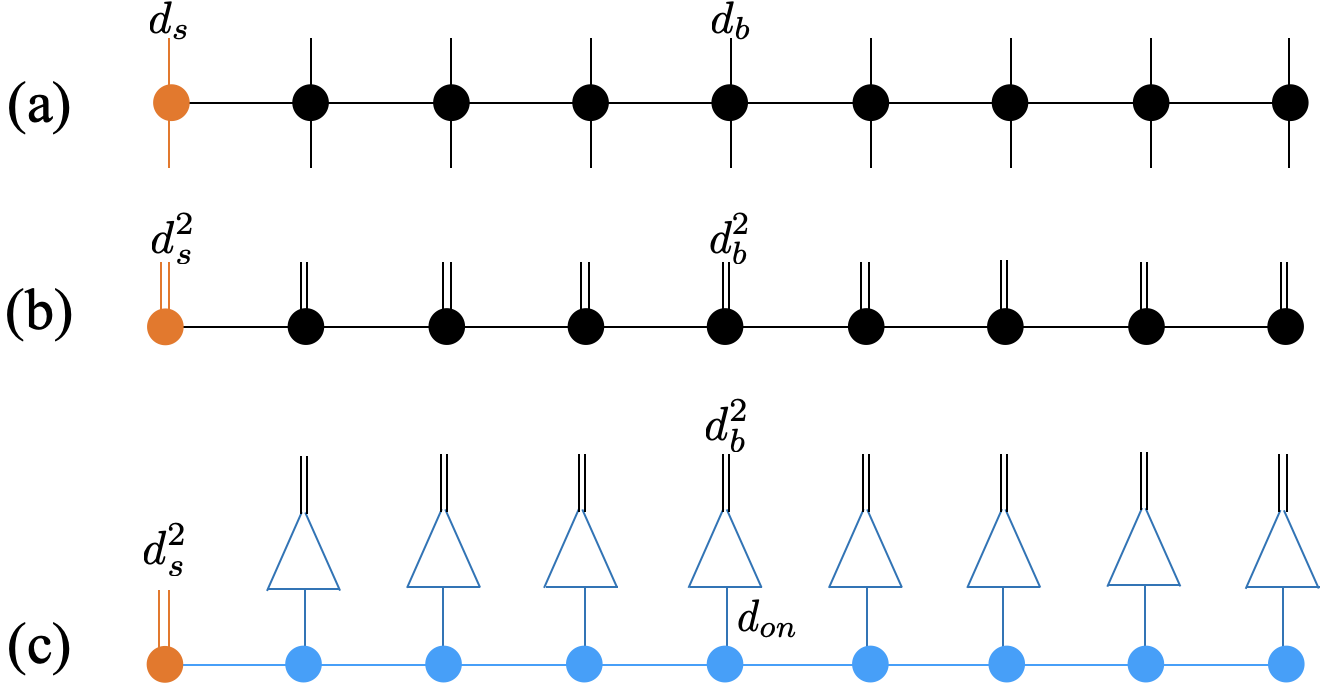}
\end{center} 
\caption{\label{fig:OBB_NAK1} Pictorial representation of the density matrix. (a) The density matrix is represented as a matrix product operator (MPO). The leftmost site represents the spin, and the remaining sites the bosonic bath. The local Hilbert space dimension of the spin and bosons are given by $d_s$ and $d_b$, respectively. (b) Vectorized form of the density matrix can be viewed as a matrix product state (MPS) by gluing bra and ket indices giving rise to the local Hilbert space dimension  $d_s^2$ and $d_b^2$ for the spin and bosonic sites, respectively. (c) We choose an optimal local basis for an efficient representation of the MPS at large $d_b$. The bosonic MPS $A[n]$ at site $n$ [black circles in part (b)]  is replaced by $\tilde{A}[n]R[n]$, where $\tilde{A}[n]$ (blue circles) is the MPS in an optimal basis and $R[n]$ (triangular box) is a unitary transformation of size $d_{on}\times d_b^2$ (see the text).
}
\end{figure}

\subsection{Optimal bosonic basis}
In the optimal bosonic basis approach, we truncate the local Hilbert space of a state \( |\rho\rangle\rangle \) at site \( n>0 \) (in the spin-boson case) using the eigenvectors of its local reduced density matrix, defined as
\begin{equation}
\rho[n] = \mathrm{Tr}_{\{1,2,\ldots,L\} \setminus n} \left[ |\rho\rangle\rangle \langle\langle \rho| \right].
\end{equation}
Assuming that the MPS tensors \( A[i] \) are in left-canonical form for \( i < n \) and right-canonical form for \( i > n \) \cite{SCHOLLWOCK201196}, the reduced density matrix \( \rho[n] \) can be expressed as
\begin{equation}
\rho[n]_{s_n,\tilde{s}_n} = \sum_{\alpha_{n-1},\alpha_n} A[n]^{s_n}_{\alpha_{n-1},\alpha_n} \left( A[n]^{\tilde{s}_n}_{\alpha_{n-1},\alpha_n} \right)^{\ast}.
\end{equation}
The eigenvectors of \( \rho[n]_{s_n,\tilde{s}_n} \) corresponding to the largest eigenvalues define the local optimal basis, with weights given by the eigenvalues. In this basis, the tensor \( A[n]^{s_n} \) can be expressed as
\begin{equation}
A[n]^{s_n} = \sum_{s_{on}=1}^{d_{on}} R_{s_n,s_{on}}[n]\, \tilde{A}[n]^{{s_{on}}},
\end{equation}
where \( R_{s_n,s_{on}}[n] \) are the eigenvectors of \( \rho[n] \), \( \tilde{A}[n]^{s_{on}} \) are the tensors in the truncated (optimal) basis, and \( d_{on} < d_b^2 \) is the optimal local Hilbert space dimension. When \( d_{on} = d_b^2 \), the transformation is exact.

With this transformation, the MPS ansatz for \( |\rho\rangle\rangle \) becomes
\begin{equation}
|\rho\rangle\rangle = \sum_{\{s_n\}, \{s_{on}\}} A[0]^{s_0}
\left( \prod_{n=1}^{L} R_{s_n, s_{on}}[n]\, \tilde{A}[n]^{s_{on}} \right) 
|s_0,s_1, s_2, \ldots, s_L\rangle.
\end{equation} 
The evolution of  \( R_{s_n, s_{on}} \) and \( \tilde{A}[n]^{s_{on}} \) using TEBD is described in detail in Refs.~\cite{Brockt_Optimal_Bosonic_Basis,Stolpp_OBB}. The index corresponding to the optimal basis is denoted by prefixing it with \( o \).

A pictorial representation of the density matrix in matrix product operator form, its vectorized representation, and the optimal bosonic basis is depicted in Fig.~\ref{fig:OBB_NAK1}.

\subsection{Further results on phase transition using MPS}

 In this section, we present numerical results obtained using time-evolving block decimation (TEBD) combined with an optimal bosonic basis; see the previous section. We start with an initial state where the spin is in \(|+\rangle\) state while the bosons are in their vacuum state.
In our numerical calculation, we set $r=1/2$, $\omega_c=10$, $\Delta=1$, and system size $L=50$. We dynamically update both the MPS bond dimension $\chi$ and the optimal local Hilbert space dimension $d_{\text{on}}$. The maximum bond dimension is fixed at $\chi = 50$, while $d_{\text{on}}$ can reach values between 20 and 30 for $d_b = 12$.

\begin{figure}[h!]
\begin{center}
\includegraphics [ scale=0.4]
{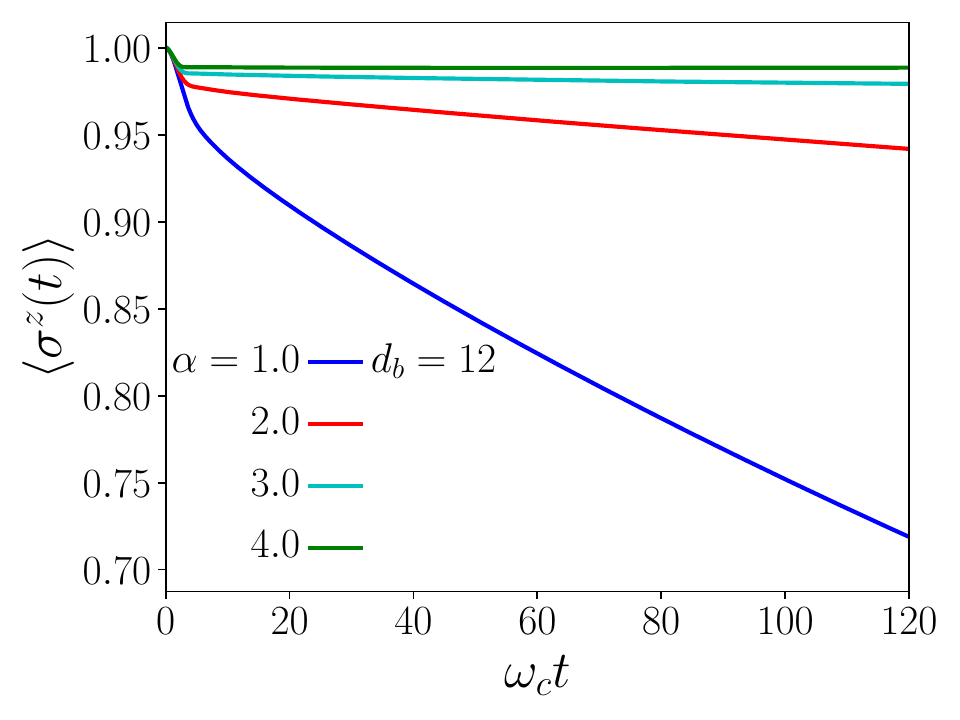}
\includegraphics [ scale=0.4]
{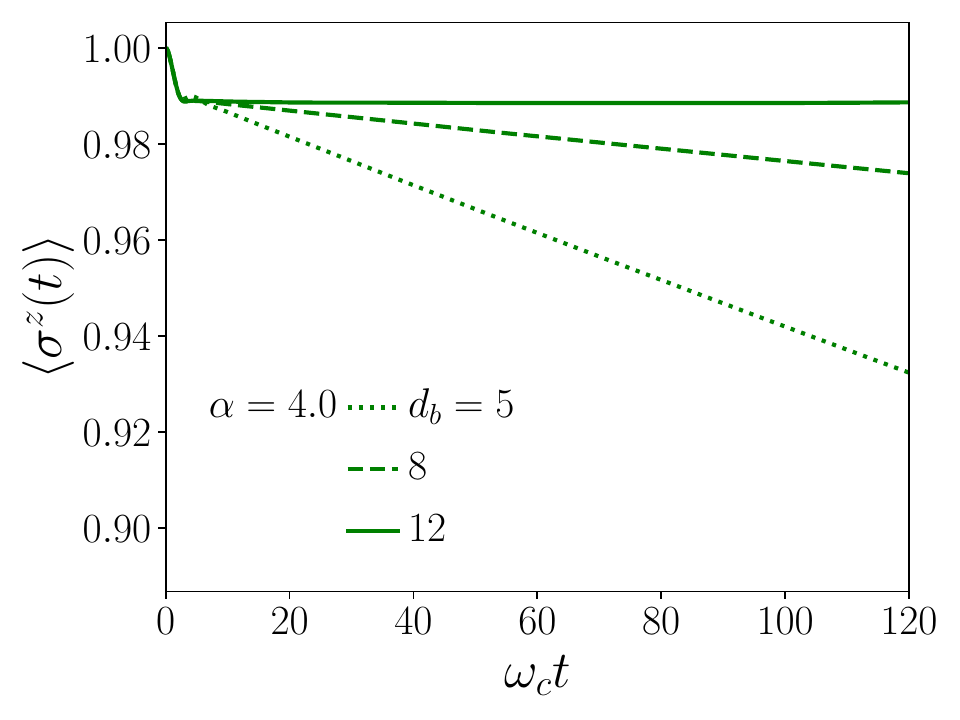}
\end{center} 
\caption{\label{fig:fig1_NAK} The magnetization dynamics, $\langle
\sigma^z(t)\rangle$ as a function of time with dissipation acting on all sites; we set $r=1/2$, $\omega_c=10$, $\Delta=1$. Left Panel: For a fixed local Hilbert space dimension of bosonic modes $d_b=12$, we observe a transition at $\alpha=4$ where $\langle \sigma^z(t) \rangle$ rapidly approaches a stationary value close to 1. Right Panel: The magnetization $\langle \sigma^z(t)\rangle$ is plotted for $\alpha=4$ at different values of $d_b=5,8,12$. As $d_b$ increases, the magnetization approaches a constant; the phase transition is observed only at the largest $d_b$, where the magnetization no longer decays.
}
\end{figure}

In \cref{fig:fig1_NAK} (left panel), we show the time evolution of the magnetization \(\langle \sigma^z(t) \rangle\) with dissipation on all sites (corresponding to the original spin-boson model) and for different values of \(\alpha=1,2,3,4\). We find that at large coupling, \(\alpha = 4.0\), the magnetization \(\langle \sigma^z(t) \rangle\) quickly reaches a stationary value close to 1, indicating a quantum phase transition to a localized phase. To verify convergence with $d_b$, we further examine \(\langle \sigma^z(t) \rangle\) at \(\alpha = 4.0\) for increasing local Hilbert space dimensions \( d_b= 5, 8,12 \). The results show a strong dependence on \( d_b \); however, the magnetization appears to converge with increasing \( d_b \). We observe that a true phase transition occurs at the largest value of $d_b$. It is possible that more intensive numerics with further increasing $d_b$ shows a phase transition at a smaller value of $\alpha$.

\begin{figure}[h!]
\begin{center}
\includegraphics [ scale=0.4]
{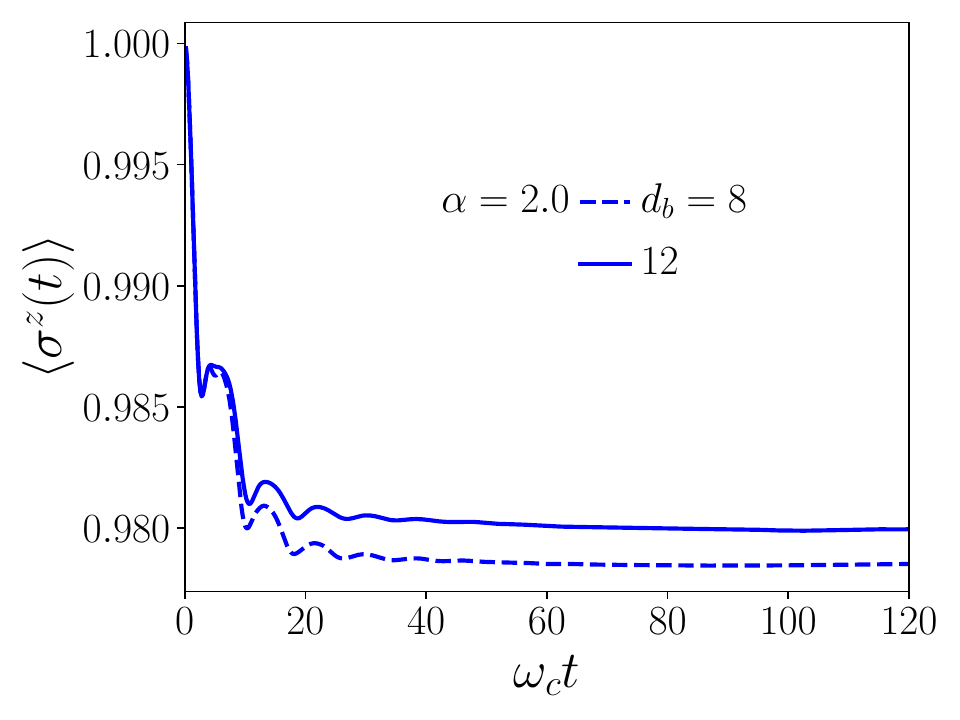}
\end{center} 
\caption{\label{fig:fig2_NAK} The magnetization dynamics, $\langle \sigma^z(t) \rangle$, with dissipation on all sites except $n=0$; parameters as in \cref{fig:fig1_NAK}, with $\alpha = 2$ and two values of $d_b = 8, 12$. Numerical results are nearly converged, with magnetization tending slightly closer to 1 as $d_b$ increases.
}
\end{figure}

In \cref{fig:fig2_NAK}, we plot the dynamics of the tweaked model with dissipation on all bosonic sites except $n=0$. In this case, we observe a phase transition already at \( \alpha = 2 \) and even with $d_b=8$. Furthermore, the numerics is converged with $d_b$;  increasing the local Hilbert space dimension to $d_b=12$  only slightly pushes the magnetization closer towards $\alpha=1$.

\section{Large Spin Coupled to Bosons: Quantum vs Classical}
\label{sec:large_spin}

In this section, we study an alternative model where the bosonic bath is coupled to a large spin, rather than a single spin $1/2$. The full Hamiltonian of the system in the original basis is given by
\begin{equation}
    H = \Delta S_z + \sum_{k=0}^{L-1} \omega_k a_k^\dag a_k + \frac{S_x}{\sqrt{N}} \sum_{k=0}^{L-1} \lambda_k (a_k^\dag + a_k),
\end{equation}
where $N$ is the total number of spins. Compared to \magenta{Eq.~(2)} in the main text, we have performed a $-\pi/2$ rotation around the $y-$axis, which takes $S_z \to S_x$, and $S_x \to - S_z$.

For $N\gg 1$, we can make a Holstein-Primakoff transformation such that 
\begin{equation}
        S_+ =S_-^\dagger \approx b^\dag \sqrt{N}, \qquad S_x=(S_+ + S_-)/2,
        \qquad S_z = b^\dag b -N/2, 
\end{equation}
where $b, b^\dag$ are bosonic annihilation and creations operators (not to be confused with the bosonic operators in the lattice representation). We then arrive at a Hamiltonian quadratic in the bosonic operators 
\begin{equation}
    H = \Delta b^\dag b +\sum_{k=0}^{L-1} \omega_k a_k^\dag a_k + (b^\dag +b )\sum_{k=0}^{L-1}\frac{\lambda_k}{2}  (a_k^\dag +a_k).
\end{equation}
The full dynamics follow the Lindblad master equation
\begin{equation}
    \dv{}{t}\rho = - i \comm{H}{\rho} + \sum_{k=0}^{L-1}2 L_k \rho L_k^\dag - \acomm{L_k^\dag L_k}{\rho},
    \label{eq:lindblad}
\end{equation}
where the jump operators are given by $L_k = \sqrt{\kappa_k} a_k$ with $\kappa_k=r \omega_k$.

\subsection{Fluctuations}
To study this model, we utilize standard field-theory techniques \cite{DallaTorre13,Sieberer2016,Lundgren_2020} to integrate out the bath modes (i.e., the fields $a_k$) arriving at the effective Keldysh action
\begin{equation}
    \mathcal{S}[\{ b, b^* \}] = \int_\omega v^\dag(\omega) ~\begin{pmatrix}
        0 & \boldsymbol{P}^A(\omega) \\
        \boldsymbol{P}^R(\omega) & \boldsymbol{P}^K(\omega)
    \end{pmatrix}~ v(\omega),
\end{equation}
where $v(\omega) = (b_c(\omega) , ~ b_c^*(-\omega),~ b_q(\omega),~b_q^*(-\omega))^T$, and $\int_\omega = \int_0^\infty d\omega/2\pi$, with $b_{c/q}$ the classical/quantum Keldysh fields corresponding to the operator $b$. The inverse retarded and advanced Green's functions are given by
\begin{equation}
    \boldsymbol{P}^R(\omega) = [\boldsymbol{P}^A(\omega)]^\dag = \begin{pmatrix}
        \Sigma(\omega) +\omega - \Delta & \Sigma(\omega) \\
        \Sigma(\omega) & \Sigma(\omega) -\omega -\Delta
    \end{pmatrix},
\end{equation}
and the Keldysh component 
\begin{equation}
    \boldsymbol{P}^K(\omega) = d(\omega) ~ \begin{pmatrix}
        1 &1 \\
        1& 1
    \end{pmatrix}.
\end{equation}
where the self energy function $\Sigma(\omega)$ and the function $d(\omega)$  are defined in terms of the bath spectral function as
\begin{subequations}
    \begin{align}
        &\Sigma(\omega)= -\frac{1}{2} \int_{\Tilde{\omega}} J(\Tilde{\omega}) \Big[ \frac{1}{(\omega-\Tilde \omega) +i \kappa(\Tilde \omega)} -\frac{1}{(\omega+\Tilde \omega) +i \kappa(\Tilde \omega)}  \Big], \\
        &d(\omega) =i \int_{\Tilde{\omega}} \kappa(\Tilde \omega)J(\Tilde{\omega}) \Big[ \frac{1}{(\omega-\Tilde \omega)^2 + \kappa^2(\Tilde \omega)} +\frac{1}{(\omega+\Tilde \omega)^2 + \kappa^2(\Tilde \omega)}  \Big].
    \end{align}
\end{subequations}

\begin{figure}[t!]
    \centering
    \includegraphics[width=0.35\linewidth]{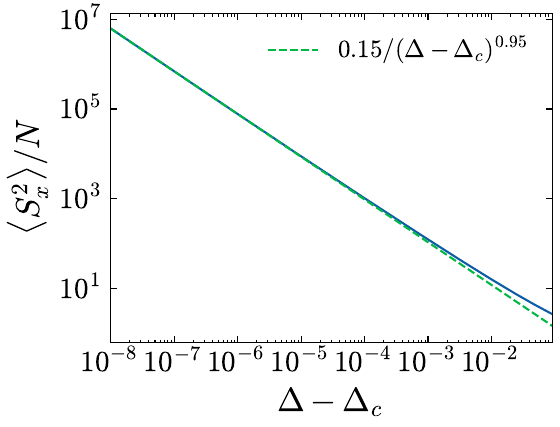}
    \caption{Spin fluctuations near the critical point $\Delta_c$. Here, we take $\alpha= 1,~ r = 1/2, ~\omega_c=10$. $\expval{S_x^2}/N$ is calculated numerically from \cref{eq:n} showing fluctuations to follow a classical phase transition scaling $\expval{S_x^2}/N \sim 1/(\Delta-\Delta_c)$.}
    \label{fig:crit_fluctuations}
\end{figure}

A superradiant phase transition occurs when $\det \boldsymbol{P}^R(\omega=0, \Delta=\Delta_c) =0$, or
\begin{equation}
    \begin{split}
        \Delta_c &= 2 \Sigma(\omega=0)=\frac{2 \alpha \omega_c}{1+r^2}.
    \end{split}
\end{equation}
Unlike the spin-boson model,
a continuous phase transition occurs for any $\alpha$ when $\Delta = \Delta_c \propto \alpha$. Furthermore, this transition is simply a superradiant phase transition where $N\to \infty$ plays the role of the thermodynamic limit. Here, we are  interested in the critical nature of the phase transition.
To understand the critical behavior, we first break the fields into their real and imaginary parts by defining
\begin{equation}
    \phi_{c/q}(t) = 2 \Re  b_{c/q}(t), ~~~~~~~ \zeta_{c/q}(t) = 2 \Im  b_{c/q}(t).
\end{equation}
One can see that the fields $\zeta_{c/q}$ are gapped, and we can integrate them out close to the critical point. We then find an effective Keldysh action depending only on $\phi_{c,q}$ as
\begin{equation}
    \mathcal{S}[\phi_{c/q}] = \int_\omega \begin{pmatrix}
        \phi_c(\omega) \\
        \phi_q(\omega)
    \end{pmatrix}^\dag ~\begin{pmatrix}
        0 & P^A_\phi(\omega) \\
         P^R_\phi(\omega) &  P^K_\phi(\omega)
    \end{pmatrix}~\begin{pmatrix}
        \phi_c(\omega) \\
        \phi_q(\omega)
    \end{pmatrix},
\end{equation}
where
\begin{subequations}
    \begin{align}
        &P^R_\phi (\omega) = [P^A_\phi(\omega)]^* = \Sigma(\omega) - \frac{\Delta}{2} +\frac{\omega^2}{2\Delta},\\
        &P^K_\phi(\omega) = d(\omega) .
    \end{align}
\end{subequations}
The large spin excitations are captured by the quantity
\begin{equation}\label{eq:n}
        \frac{\expval{S_x^2}}{N}= \frac{1}{4}  \int_\omega \expval{\phi_c^2(\omega)}
        = -\frac{i}{4} \int_\omega \frac{P^K_\phi (\omega)}{\abs{P^R_\phi}^2}.
\end{equation}
Approaching the critical point ($\Delta \to \Delta_c$), the low-frequency contribution diverges as $P^R_\phi(\omega \to 0)$ vanishes, while $P^K_\phi(\omega \to 0)$ remains finite. The integral in \cref{eq:n} is calculated numerically and shown in \cref{fig:crit_fluctuations}, where we find that the spin excitations diverge as $1/(\Delta -\Delta_c)$ upon approaching the critical point. This scaling is compatible with classical rather than quantum fluctuations; see the main text. We find consistent results for the critical fluctuations when dissipation is turned off on the site $n=0$ (see \cref{subsec:L_0=0}).

\subsection{Purity}
To further highlight the classical nature of the phase transition, we calculate the quantum purity of the steady state as a function of the distance from the critical point ($\Delta-\Delta_c$).  In this subsection, it is more convenient to work in the lattice representation with the large spin (represented by the operators $b,b^\dagger$) coupled to a bosonic lattice of size $L$. We denote the full set of bosons by the vector $\underline{a}=(a_0,...,a_{L-1},b)$. 

For a Gaussian system, the purity can be computed as \cite{illuminati2004}
\begin{equation}
    \mu = \frac{1}{\sqrt{\det \sigma}}.
\end{equation}
where the covariance matrix $\sigma$ is defined as
\begin{equation}
    \sigma = \begin{pmatrix}
        \expval{\acomm{\underline{x} }{\underline{x} }} & \expval{\acomm{\underline{x} }{ \underline{p} }} \\
        \expval{\acomm{\underline{p}}{ \underline{x} }} & \expval{\acomm{\underline{p} }{ \underline{p} }}
    \end{pmatrix},
\end{equation}
with the expectation values taken with respect to the steady state. The vectors $\underline{x}$ and $\underline{p}$ represent the position and momentum operators of the bosonic modes,
\begin{subequations}
    \begin{align}
        &\underline{x}_i = \frac{1}{\sqrt{2\omega_i}} (\underline{a}_i^\dag + \underline{a}_i),\\
        &\underline{p}_i = i \sqrt{\frac{\omega_i}{2}} (\underline{a}^\dag_i -\underline{a}_i).
    \end{align}
\end{subequations}
The covariance matrix $\sigma$ can be calculated exactly using the third quantization method, similar to \cref{subsec:L_0=0}. Here, we expand the Hamiltonian matrix to include the coupling of the chain to the large spin (now mapped to a soft-core boson) and define
\begin{equation}
    \mathbf{H}' = 
\left(
\begin{array}{cccc|c}
  &        &    &   & -i c_0 \\
  &        &    &   & 0 \\
  &  & \mathbf{H} &  & \vdots \\
  &        &    &   & 0 \\
\hline
i c_0 &0& \cdots & 0 & \Delta
\end{array}
\right).
\end{equation}
Similarly, we expand the matrix $\mathbf{M}$ to an $(L+1) \times (L+1)$ matrix
\begin{equation}
    \mathbf{M}' = 
\left(
\begin{array}{ccc|c}
  &        &       & 0 \\
  & \mathbf{M} &   & \vdots \\
  &        &       & 0 \\
\hline
0 & \cdots & 0 & 0
\end{array}
\right),
\end{equation}
that accounts for the dissipation terms, and define $\mathbf{X}' = \frac{1}{2} \begin{pmatrix}
        i (\mathbf{H}')^* +\mathbf{M}' & 0 \\
        0 & - i \mathbf{H}' + (\mathbf{M}')^*
    \end{pmatrix}$.
    
\begin{figure}[t!]
    \centering
    \includegraphics[width=0.35\linewidth]{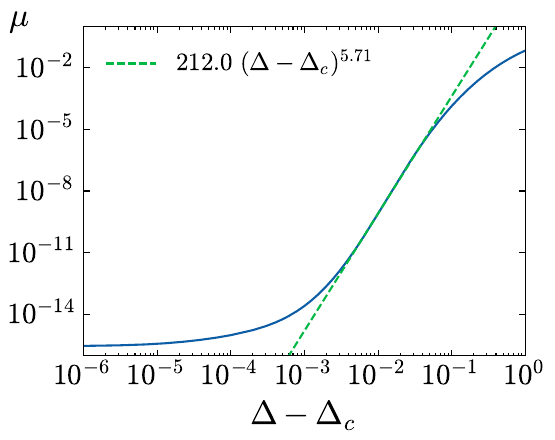}
    \caption{Purity in the steady state. We set the parameters $\alpha= 1,~ r = 1/2, ~\omega_c=10$, and consider a bosonic bath of size $L=1000$. Purity quickly vanishes upon approaching the critical point, following a power law in $\Delta - \Delta_c$ with an exponent of approximately 6, and levels off very close to the critical point due to finite-size effects.}
    \label{fig:purity_plot}
\end{figure}

Solving the Lyapunov equation \cite{seligman2010}
\begin{equation}
    \mathbf{X}'~\mathbf{Z}+\mathbf{Z}~\mathbf{X}'=0,
\end{equation}
allows us to compute the matrix
\begin{equation}
    \mathbf{Z} = \begin{pmatrix}
        \expval{\underline{a} \otimes \underline{a} } & \expval{:\underline{a} \otimes \underline{a}^\dag :} \\
        \expval{\underline{a}^\dag \otimes \underline{a} } & \expval{\underline{a}^\dag \otimes \underline{a}^\dag }
    \end{pmatrix} = \begin{pmatrix}
        \mathbf{Z}_{11} & \mathbf{Z}_{12}\\
        \mathbf{Z}_{21} & \mathbf{Z}_{22}
    \end{pmatrix}.
\end{equation}
The elements of the covariance matrix can then be computed by noting
\begin{subequations}
    \begin{align}
        &\expval{\acomm{\underline{x} }{\underline{x} }}_{ij} =2\expval{x_ix_j}= \frac{1}{\sqrt{\omega_i \omega_j}} (\mathbf{Z}_{11} + \mathbf{Z}_{12} + \mathbf{Z}_{21} + \mathbf{Z}_{22} + \mathbf I)_{ij},\\
        &\expval{\acomm{\underline{p} }{\underline{p} }}_{ij} =2\expval{p_ip_j}= \sqrt{\omega_i \omega_j} (-\mathbf{Z}_{11} + \mathbf{Z}_{12} + \mathbf{Z}_{21} - \mathbf{Z}_{22} + \mathbf I)_{ij},\\
        &\expval{\acomm{\underline{x} }{\underline{p} }}_{ij} =\expval{x_ip_j+p_jx_i}= i \sqrt{\frac{\omega_j}{\omega_i}} (-\mathbf{Z}_{11} + \mathbf{Z}_{12} - \mathbf{Z}_{21} + \mathbf{Z}_{22})_{ij},\\
        &\expval{\acomm{\underline{p} }{\underline{x} }}_{ij} = \expval{\acomm{\underline{x} }{\underline{p} }}_{ji}.
    \end{align}
\end{subequations}
We can now calculate the purity efficiently from the covariance matrix by numerically computing various correlation functions. 
We focus on the behavior of the purity near the critical point. In \cref{fig:purity_plot}, we plot the purity as a function of the distance from the critical point; we set the parameters $\alpha=1,~r=1/2,~\omega_c= 10$ (the critical point is $\Delta_c = 16$). We find that the purity quickly vanishes upon approaching the critical point from the normal phase, further underscoring the classical nature of the transition.

\bibliographystyle{apsrev4-2}